\documentclass[review]{elsarticle}

\usepackage{soul}
\usepackage{amsmath}
\usepackage{amssymb}

\usepackage{geometry}
\usepackage{setspace}
\usepackage{textcomp}
\usepackage{parskip}
\usepackage{placeins}
\usepackage{float}
\usepackage{multirow}
\usepackage[figurename=Fig.]{caption}
\usepackage{array}
\usepackage{svg}
\usepackage{pdfpages}
\usepackage{subcaption}

\usepackage{soul}
\usepackage{color}

\usepackage{gensymb}
\usepackage{siunitx}

\begin{document}

\begin{frontmatter}

%\title{Elsevier \LaTeX\ template\tnoteref{mytitlenote}}
\title{\textbf{A Surrogate-model-based Approach for Estimating the First and Second-order Moments of Offshore Wind Power}} %Prediction}}


%\author[mymainaddress,mysecondaryaddress]{first author}

\author[mymainaddress]{Behzad Golparvar}
\author[mysecondaryaddress]{Petros Papadopoulos}
\author[mysecondaryaddress]{Ahmed Aziz Ezzat}%\corref{mycorrespondingauthor}}
\author[mymainaddress]{Ruo-Qian Wang\corref{mycorrespondingauthor}}
%\cortext[mycorrespondingauthor]{Co-corresponding author}
\cortext[mycorrespondingauthor]{Corresponding author}
\ead{%aa2085@soe.rutgers.edu (Ahmed Aziz Ezzat),
rq.wang@rutgers.edu}

\address[mymainaddress]{Department of Civil and Environmental Engineering, Rutgers University, Piscataway, NJ, USA}
\address[mysecondaryaddress]{Department of Industrial and Systems Engineering, Rutgers University, Piscataway, NJ, USA}

\begin{abstract}
Power curve, the functional relationship that governs the process of converting a set of weather variables experienced by a wind turbine into electric power, is widely used in the wind industry to estimate power output for planning and operational purposes. Existing methods for power curve estimation have three main limitations: (i) they mostly rely on wind speed as the sole input, thus ignoring the secondary, yet possibly significant effects of other environmental factors, (ii) they largely overlook the complex marine environment in which offshore turbines operate, potentially compromising their value in offshore wind energy applications, and (ii) they solely focus on the first-order properties of wind power, with little (or null) information about the variation around the mean behavior, which is important for ensuring reliable grid integration, asset health monitoring, and energy storage, among others. 
 %Existing methods for constructing power curves mostly overlook the complex marine environment in which offshore wind turbines operate, thus compromising their value in offshore wind energy applications and the power output fluctuation that is pertaining to grid integration and energy storage.
In light of that, this study investigates the impact of several wind- and wave-related factors on offshore wind power variability, with the ultimate goal of accurately predicting its first two moments. Our approach couples OpenFAST\textemdash a high-fidelity stochastic multi-physics simulator\textemdash with Gaussian Process (GP) regression to reveal the underlying relationships governing offshore weather-to-power conversion. We first find that a multi-input power curve which captures the combined impact of wind speed, direction, and air density, can provide double-digit improvements, in terms of prediction accuracy, relative to univariate methods which rely on wind speed as the sole explanatory variable (e.g. the standard method of bins). Wave-related variables are found not important for predicting the average power output, but interestingly, appear to be extremely relevant in describing the fluctuation of the offshore power around its mean. Tested on real-world data collected at the New York/New Jersey bight, our proposed multi-input models demonstrate a high explanatory power in predicting the first two moments of offshore wind generation, testifying their potential value to the offshore wind industry. 
\end{abstract}

\begin{keyword}
Offshore wind\sep Multi-input power curve\sep Power uncertainty\sep OpenFAST\sep Machine learning
%\MSC[2010] 00-01\sep  99-00
\end{keyword}

\end{frontmatter}

\begin{table}
\centering
\resizebox{\columnwidth}{!}{\begin{tabular}{|llllll|} 
\hline
            &                                           &  &                    &            &                                       \\
\multicolumn{6}{|l|}{\textbf{Nomenclature} }                                                                                        \\
            &                                           &  &                    &           &                                        \\
$A$         & Area of Turbine Rotor [m$^2$]           &  & \multicolumn{3}{l|}{\textit{Greek symbols} }                           \\
$C_p$       & Power Coefficient [-]                     &  &                    &      &                                             \\
$H$         & Wave Height [m]                           &  & $\beta$            & Wave Direction [deg]  &                            \\
$P$         & Generator Power [kW]                      &  & $\theta$           & Relative Wind Direction/Yaw Misalignment [deg] &   \\
$P_{m}$     & Mean Generator Power [kW]                 &  & $\rho$             & Air Density [m$^3$/kg]   &                         \\
$P_{sd}$    & Generator Power Standard Deviation [kW]   &  & $\phi$             & Relative Humidity [\%]   &                         \\
$P_{a}$     & Partial Pressure of Dry Air [Pa]          &  &                    &         &                                          \\
$P_{sat}$   & Pressure of Water Vapor [Pa]              &  & \multicolumn{3}{l|}{\textit{Abbreviations} }                            \\
$R_{d}$     & Specific Gas Constant for Dry Air [J/kg.K]        &  &                    &      &                                  \\
$R_{v}$     & Specific Gas Constant for Water Vapor [J/kg.K]    &  & GP                 & Gaussian Process     &                   \\
$T$         & Temperature [\textdegree{}C]              &  & MAE                & Mean Absolute Error                  &             \\
$V$         & Wind Speed [m/s]                          &  & NRMSE              & Normalized Root Mean Squared Error    &            \\
            &                                           &  & RMSE               & Root Mean Square Error               &             \\

\hline
\end{tabular}}
\end{table}

\section{Introduction}

%\subsection{Significance of wind energy:}
Offshore wind is rapidly maturing into a major source of renewable energy, which is projected to grow by $13$\% in the next two decades and fifteen-fold by 2040 to become a \$$1$ trillion industry, matching the capital spending on gas- and coal-fired power generation \cite{Offshore94online}. 
In the U.S., for instance, New York and New Jersey have recently awarded two offshore wind energy contracts, in total of $3$ GW, to achieve their targets of renewable energy integration. Similar, and even more ambitious projects are under planning \cite{NewYorkT16:online}. This progress unveils a new age of offshore wind energy revolution. The key to support this offshore wind energy growth is to develop reliable tools to assess, and further predict, offshore wind energy potential in order to effectively inform project planning, and support offshore operations and maintenance. On the other hand, a growing concern about the reliability of renewable-dominant electricity systems, especially after the 2019 Hornsea offshore wind farm blackout in England \cite{National73:online} and 2021 Texas Power Crisis \cite{TexasPow31:online}, motivates an urgent need to develop powerful models to estimate/predict the environmental uncertainty of wind power generation.

Perhaps the most common tool to estimate power generation is the so-called ``power curve'', which is the functional relationship that 
governs the process of converting a set of weather variables experienced by a wind turbine into electric power. Each turbine model comes with a \textit{theoretical power curve}, also known as the manufacturer's power curve or nominal power curve, which is often attained under idealistic operational and environmental conditions. In practice, such idealistic conditions are seldom realized, and hence, \textit{actual power curves} often depart from theoretical power curves due to a combination of environmental and operational factors. It is thus important to statistically construct actual power curves by leveraging the turbine-specific weather and power data. % measurements collected using hub-height sensors mounted on a wind turbine during its operation. 
For wind farm developers and operators, an accurate estimate of a power curve is of vital importance, owing to its relevance to several critical planning and operational decisions including but not limited to turbine productivity and efficiency assessment \citep{niu2018evaluation}, asset health monitoring and prognostics \citep{gill2011wind}, power output assessment \cite{ouyang2017modeling}, forecasting \cite{ezzat2019spatio}, and optimization \citep{gebraad2017maximization}, turbine control and wake steering \citep{howland2019wind}, maintenance scheduling \citep{pap2020}, among others. 

The current industrial standard for estimating actual power curves is through the \textit{method of bins}, also known as the \textit{binning method} \citep{iec2005}. The essence of the binning method is to divide the wind speed domain into a number of bins, and then take the average power output within each bin as the estimated power output. While simple and practical, the binning method suffers from a number of limitations. First, it \textit{only} takes the wind speed as input. While wind speed is the major determinant of wind power, other environmental and operational variables may have secondary, yet significant effects on the turbine power output. In fact, this is evident by the physical law governing the wind power generation expressed as $P = \frac{1}{2} C_p \rho A V^3$, where $P$ is the power output, $C_p$ is the power coefficient, $\rho$ is the air density, $A$ is the area of turbine's rotor, and $V$ is the wind speed. This physical law suggests that wind speed is not the sole determinant of wind power, albeit being the most important. Other factors such as air density, which, in turn, is governed by the values of temperature and pressure, may play an important role. In addition, the power coefficient, $C_p$ is, in itself, a function of the turbine's design and operational parameters. \par  

Second, %the traditional power curve is deterministic, i.e. it does not associate the estimated power curve with probabilistic information such as the second-order effect of power generation. 
the method of bins is designed to solely describe the ``average'' behavior or first-order properties of wind power, but fails, in its standard form, to provide information about higher-order moments, e.g., the variability around the mean behavior. A turbine system, including the rotor, nacelle, tower, foundations, etc., is a dynamic system that exhibits continuous vibration due to the periodic movements of its connected components. Such vibrations may cause the power output to exhibit sizeable fluctuations, which in turn can have serious implications on the compatibility and stability of power storage and grid integration. %Moreover, vibration is an important factor in determining the life span of wind turbine systems because the amplitude and frequency of the vibration dictates the fatigue acting on its critical components. 
These second-moment properties, i.e., the power output variation around the mean behavior, are believed to be more relevant in offshore than in onshore wind farms due to the compounded impact of marine conditions, from one hand, and to the increasing scale of offshore wind turbines from the other. %In fact, our numerical results interestingly show that wave-related variables are strongly correlated with the second-order properties of offshore wind power. 

%However, the traditional statistics based method has difficulty to estimate this secondary effect as such information is missing in the ``power curve''. \hlc[pink]{Aziz Task1: standardize the term use.}\hl{[Comment by Aziz: I feel we need to standardize what we use to refer to the ``variance'' in power curve: In this paragraph alone, we used both second order effect and secondary effect. Are we sure that ``second order effect'' is clearly understood as ``variance''? For instance, in statistics/design of experiments, a second-order effect is usually used to refer to the impact of a combination (interaction) of two variables on the response]. Roger: have we addressed the issue?}

Third, most existing studies are targeted at estimating power curves in onshore settings. Offshore wind turbines operate in a different and less understood marine environment compared to their onshore counterparts, so the marine environmental factors are rarely considered. For instance, it is not yet clear what the offshore-specific variables such as wave height, direction, and air-sea interactions would impact on the first- and second-order properties of offshore wind power. %With the additional uncertainties of these offshore-specific variables, and with the unprecedented altitudes and dimensions of modern offshore turbines, there is a need to establish statistical methods that rigorously and collectively accommodate the impact of these offshore-specific features.
Up to our knowledge, there is no offshore-specific power curve estimation methodology which systematically examines the effect of environmental and operational uncertainties in offshore settings. The present paper is targeted to fill this gap by proposing a data-driven offshore-specific power curve estimation method which accurately reconstructs the multi-input weather-to-power conversion in an offshore setting, hence facilitating its application to the growing offshore-specific industry. \par  

Motivated by the aforementioned observations, an active line of research in recent years is concerned with proposing data-driven methods to construct what we call hereinafter as \textit{multi-input power curves}, i.e., power curves that, in addition to wind speed, take into account other environmental and operational factors in determining a turbine's power output. For instance, incorporating hub-height air density measurements using a multi-input Gaussian Process has been recently shown to benefit wind power curve estimation \citep{pandit2019incorporating}. A bivariate kernel density estimation based on wind speed and direction has also been shown to yield significant improvements in prediction accuracy relative to the binning method \citep{jeon2012using}. Other factors such as air density, turbulence intensity, and wind shear were also proven to be useful for estimating multi-input power curves \citep{lee2015power}. Overall, these methods have successfully tackled the first limitation of the binning method mentioned above, i.e. the inability to account for several environmental variables, but not the other two, i.e., characterization of higher-order moments, and amenability to offshore environments. \par

To support the development of the multi-input power curve, we need turbine-specific data. There are potentially two methods to acquire such data: field observation and numerical simulation. In this study, we focus on the latter approach, which enables us to construct a large and diverse database of \textit{all} relevant weather variables and associated power generation, following a planned experimental design allowing for wider exploratory coverage of the input-output relationship. This is in contrast with field observations that are often incomplete in variety and limited in frequency and coverage and difficult to access due to technical or security reasons. In particular, we make use of OpenFAST, which is a high-fidelity, open-source wind turbine simulation tool developed by the National Renewable Energy Laboratory (NREL) via support from the United States Department of Energy \cite{jonkman2018full}. OpenFAST is a multi-physics simulator, i.e., it couples the aero-hydro-servo-elastic sub-models to simulate the time-domain physical interaction among the wind turbine, environment, and control systems \cite{jonkman2018full}. In addition, OpenFAST is able to conduct the structure vibration analysis, support control system design, analyze structure stability, and provide gradients for optimization \cite{johnson2019verification}. To date OpenFAST (and its earlier version of ``FAST'') has been used to design offshore floating wind turbines \cite{jonkman2018full}, estimate wind energy resources \cite{wang2020multi}, analyze wake effects \cite{wise2019analysis}, test new generator compatibility \cite{fadaeinedjad2008simulation}, optimize wind farm layouts \cite{tran2013fast} and turbine control systems \cite{shariatpanah2013new}. %We stress, however, that our proposed models, albeit ``trained'' on numerical simulations, are tested on real-world offshore measurements collected from the New York/New Jersey Bight, where several offshore wind projects are already in the active development\textemdash phase (see more details in Section 3.4).  
Because OpenFAST allows us to perform a comprehensive study to explore the impact of several wind- and wave-related variables on offshore wind power, we employ it to model offshore wind energy systems to inform the sensitivity of the power curve to a variety of environmental factors. %Second, because of its rich results in the dynamic response of the wind turbine to the environments, we can obtain the first and second-order effects of power generation and structure stability. Third, this model can simulate the wind turbine that has yet released to the market, which supports future planning tasks.
%are shown to be extremely useful in characterizing the first- and second-order moments of power generation and structural stability. %To realize this, we couple OpenFAST with Gaussian Process (GP) regression, a nonparametric statistical approach, which is well known for its capability to characterize complex response surfaces emanating from physics-based simulators \cite{santner2003design}. %Another advantage of using OpenFAST is its ability to simulate futuristic turbine designs (up to $15$MW in capacity). Due to its high efficiency, Those ultra-scale turbines will be actively deployed in the next decade, and hence, such analysis can be vital for future planning tasks. %develop the multi-input power curve and quantify the environmental uncertainty. First, the valuable details of the system dynamics provided by this model can be used to model offshore wind energy systems to inform the sensitivity of the power curve to a variety of environmental factors. Second, because of its rich results in the dynamic response of the wind turbine to the environments, we can obtain the first and second-order effects of power generation and structure stability. Third, this model can simulate the wind turbine that has yet released to the market, which supports future planning tasks. %The capacity of offshore wind turbine design is increasing every year, developers must keep reformulating their blueprints for the new technology advance. OpenFAST provides the necessary technical details to meet this demand. 
Particularly, we choose the $15$ MW offshore reference turbine developed by the IEA Wind Task 37 \cite{IEA15MW_ORWT}. Although it has not yet been deployed in any wind farm, this ultra-scale wind turbine is projected to become popular in the world due to its high efficiency.

% \hl{[Comment by Aziz: which power curve method? This transition sentence may add a bit of confusion: We discuss the power curves for about a page, and then we present OPENFAST as an ``alternative'' for power curve methods? I guess this brings us back to an ``expected'' reviewer question that we need to hedge against: Why do we rely on OPENFAST, not actual data? In my opinion, two reasons, which we briefly say but perhaps need to emphasize and mention early on: (1) actual data for a 15MW offshore turbine does not yet exist, (2) it's very hard to get a detailed analysis on what we call here first and second-order effects from turbine data. Your thoughts?] }, 

In summary, the aim of this study is to bridge the gap between the high-fidelity OpenFAST simulator and probabilistic statistical learning approaches to accurately characterize and predict the power output variability in complex offshore environments. 
A unique feature of this study is to explore the impact of seven variables, including wind- and wave-related factors, on the first- and second-order moments of offshore power generation. Up to the authors' knowledge, this constitutes the most comprehensive examination of the environmental impact on offshore wind power variability. Moreover, this is the first study to adopt the $15$ MW turbine design as the target turbine system for analysis. 

\section{Research Methods}
%\hlc[pink]{Aziz Task2 and Behzad Task 1 and Roger Task 1: complete the method section and review.} \hl{[Comment by Aziz: our methods section needs to be enriched. If I understand correctly, our methods are as follows: (1) openfast, (2) Sobol sequences, and (3) GPs. Should we provide more details about each in this Section?] Roger: Yes, please. I just put some placeholders here. More details are welcome.}
To achieve the research goals, we first design the parameters space covering the common situations in offshore wind areas, then simulate the performance of the 15 MW wind turbine using OpenFAST within the widely covered database, and finally develop surrogate models using Gaussian Processes (GPs) to achieve high and multi-input accuracy in the complex offshore environment. 

\subsection{\textbf{Input Design Generation}} \label{InputVariables}

%Monte Carlo simulation (MCS) is used to generate the input design points at which the target responses will be simulated using OpenFAST. 
%To develop a surrogate model capable of predicting turbine output power characteristics, namely mean and uncertainty of generator power, Monte Carlo simulation (MCS) method is adopted. In this regard, 
Seven environmental variables are considered in this study, namely: wind speed, relative wind direction, wave height, wave direction, air temperature, atmospheric pressure, and relative humidity. We do not consider water depth in this analysis since OpenFAST decouples the structure dynamics above and below the water surface. %To generate samples to use them as input variables the variation range of each input variable should be defined. To include extreme conditions and develop a comprehensive model wide ranges are considered. 
The ranges of the seven variables are determined by the operational and environmental conditions typical for offshore wind farms. The cut-out speed of the $15$ MW offshore reference turbine is $25$ m/s, and hence, is used as the upper limit of the wind speed variable $(V)$. The wave height $(H)$ and wave direction $(\beta)$ are limited to the ranges of $0$-$20$ m and $0$-$180^\circ$, respectively, while the air temperature $(T)$ and relative humidity $(\phi)$ are limited to $-20$\textdegree{}C to $40$\textdegree{}C, and $0$\% to $100$\%, respectively. Schlechtingen \textit{et al.} \cite{Schlech2013} reported that the hub-height air pressure can vary by up to $10$\% depending on the weather phenomena. Hence, the range of atmospheric pressure, $(P_a)$ is set to a range of 10\% from $101,325$ Pa. Combining the values of $T$, $\phi$, and $P_a$, we can obtain air density $(\rho)$ using the following equation: 
\begin{equation}\label{eq:rho}
    \rho=\frac{P_a-\phi P_{sat}}{R_d(T+273.3)}+\frac{\phi P_{sat}}{R_v(T+273.3)},
\end{equation}
where $P_a$ and $P_{sat}$ are the partial pressure of dry air and pressure of water vapor, $R_d$, and $R_v$ are specific gas constant for dry air and water vapor with the unit of $\mbox{J}\cdot\mbox{kg}^{-1}\cdot\mbox{K}^{-1}$.

We also consider relative wind direction or yaw misalignment, denoted by $\theta$, to incorporate the effect caused by the difference between absolute wind direction and the turbine's yaw position. 
%However, for the wind direction variable, it should be noted that a modern turbine, which is equipped with a yaw control system, is able to adjust the turbine rotor orientation to the wind direction. In an idealized situation, the nacelle yaw angle and incoming wind direction should be always zero ensuring that the wind turbine is generating maximum electrical energy. The difference between these angles is called yaw misalignment. In practice, wind direction is rapidly and randomly changing relative to the rotor axis and due to the limitations in yaw rotational speed the active yaw control system may have a certain angular lag \cite{en8076286}. For example, in yaw controller of DOWEC6MW turbine, the moving average of wind direction with the window size of 30 seconds was considered to monitor the yaw misalignment and when the yaw misalignment reaches 5 degree yaw actuators change the nacelle orientation with the rate of 0.3 deg/s \cite{Storey_2014}. In addition, yaw correction is also limited by mechanical constraints as the frequent start and stop of yawing lead to load fluctuations in the yaw system \cite{EKELUND2000241}. Therefore, yaw misalignment or relative wind direction $(\theta)$ is considered in the present modeling as a representative for the environmental factor of wind direction. 
As reported in the IEC 61400-1 standard, %that in the steady extreme wind model
yaw misalignment in practice varies in the range of $\pm${30}\textdegree{} \cite{IEC61400_1}. Thus, we set the range of yaw misalignment to be $0$-$30^\circ$. Using the aforementioned ranges, we implement the Sobol sequence sampling method \cite{sobol1967distribution} to generate $1000$ samples to uniformly exhaust the design parameter space. Note that the Sobol sequence method is a widely recognized space-filling sampling approach which generates quasi-random, low-discrepancy samples to uniformly ``fill'' the multi-dimensional input space \cite{li2020efficient, jia2019investigation, zhao}.
 
\subsection{\textbf{Modeling and Simulation with OpenFAST}}

 In this study, OpenFAST is employed to simulate the high-fidelity behaviors of the wind turbine in complex environments. In particular, OpenFAST is designed to simulate the complex interactions between the environment and wind turbines. It couples aerodynamics (aero) models, hydrodynamics (hydro) models, control and electrical system (servo) dynamics models, and structural (elastic) dynamics models to reproduce the nonlinear aero-hydro-servo-elastic coupling dynamics in the time domain. The aerodynamic models use wind-inflow data to resolve the rotor-wake effects and blade-element aerodynamic loads; the hydrodynamics models simulate the regular or irregular incident waves and currents to resolve the hydrostatic, radiation, diffraction, and viscous loads on the offshore substructure; the control and electrical system models simulate the controller logic, sensors, and actuators of the blade-pitch, generator-torque, nacelle-yaw, and other control devices, in addition to the generator and power-converter components of the electrical drive; and the structural-dynamics models apply the control and electrical system reactions, apply the aerodynamic and hydrodynamic loads, adds gravitational loads, and simulate the elasticity of the rotor, drivetrain, and support structure. Coupling among all models is achieved through a modular interface and coupler. More details about the models can be found in \cite{jonkman2013new}.

\begin{figure}
    \centering
    \includegraphics[height=10cm]{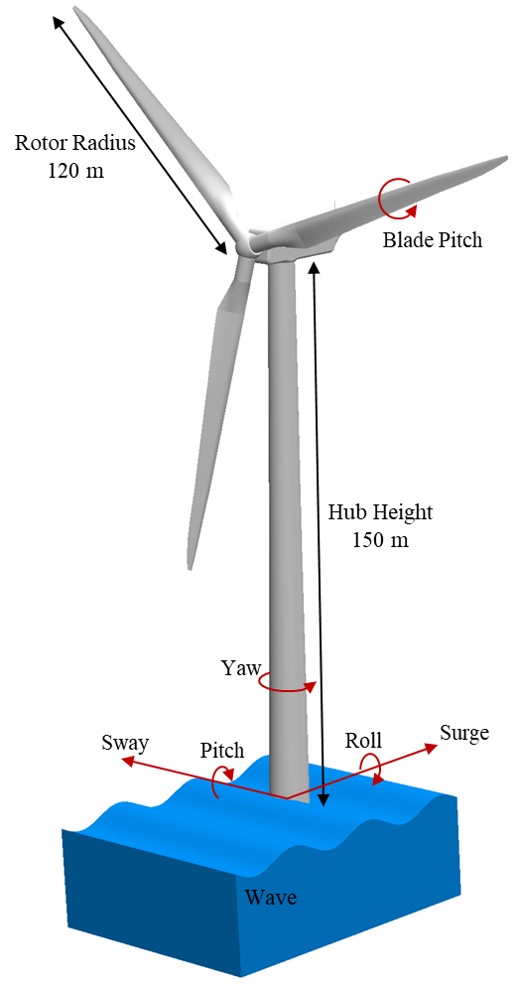}
    \caption{Rotation and movement axes, and dimensions of the IEA $15$ MW reference wind turbine in an offshore environment.}
    \label{fig:Schematic}
\end{figure}

The history of reference wind turbine models dates back to early 2000s when NREL and partners developed the $1.5$ and $3$ MW reference turbines \cite{IEA15MW_ORWT}. The $5$ MW model was added to the family and is still one of the most popular models in literature. Recently, offshore turbines of large capacities are introduced to the market such as the General Electric (GE) $12$-MW Haliade-X offshore wind turbine. This turbine has a rotor diameter of $218$ m, and is expected to be installed in 2021 in the ongoing offshore project of New Jersey in the United States. In light of that, we focus on the new wind turbine model of $15$ MW power rating released by the IEA Wind Task 3 \cite{IEA15MW_ORWT}. This reference turbine has a rotor diameter of $240$ m and a hub height of $150$ m with a monopile foundation (a semi-submersible floating platform is also available) (See Fig. \ref{fig:Schematic} for a schematic). This turbine can be seen as the future design of offshore wind turbine in the new era. The primary design parameters of this reference design are shown in Table \ref{tab:param}.

\begin{table}[]
\centering
\caption{Primary design parameters of the IEA $15$ MW reference wind turbine model}
\label{tab:param}
\begin{tabular}{lll}\hline
Parameter                   & IEA 15-MW Turbine      & Unit    \\\hline
Power Rating                & 15                     & MW      \\
Specific Rating             & 332                    & W/m$^2$ \\
Number of Blades            & 3                      & -       \\
Cut-in Wind Speed           & 3                      & m/s     \\
Rated Wind Speed            & 10.59                  & m/s     \\
Cut-out Wind Speed          & 25                     & m/s     \\
Rotor Diameter              & 240                    & m       \\
Airfoil Series              & FFA-W3                 & -       \\
Hub Height                  & 150                    & m       \\
Hub Diameter                & 7.94                   & m       \\
Hub Overhang                & 11.35                  & m       \\
Drivetrain                  & Low Speed Direct Drive & -       \\
Design Tip-Speed Ratio      & 90                     & -       \\
Minimum Rotor Speed         & 5                      & rpm     \\
Maximum Rotor Speed         & 7.56                   & rpm     \\
Maximum Tip Speed           & 95                     & m/s     \\
Blade Mass                  & 65                     & t       \\
Rotor Nacelle Assembly Mass & 1017                   & t       \\
Tower Mass                  & 860                    & t       \\
Tower Base Diameter         & 10                     & m       \\
Monopile Base Diameter      & 10                     & m       \\
Monopile Mass               & 1318                   & t       \\
Monopile Embedment Depth    & 15                     & m     \\\hline 
\end{tabular}
\end{table}

% \begin{figure}[htp]
%     \centering
%     \includegraphics[width=12cm]{total_cases_power.pdf}
%     \caption{All the healthy cases generated by OpenFAST}
%     \label{fig:total_cases_power}
% \end{figure}

\subsection{\textbf{Surrogate Modeling of Stochastic Simulators Using Gaussian Process Regression}}
 
 %\hlc[green]{Dr Aziz could you please complete this section? I am not very familiar with GP and kriging explanations}
 
 The stochastic simulation results of OpenFAST are analyzed using Gaussian process (GP) regression. GPs have shown tremendous successes in the last decade to model the outputs from complex computer simulations in physical and engineering sciences, mostly owing to their flexibility to characterize complex nonlinear response surfaces \citep{ezzat2020sur}. For example, they have been used as surrogates to approximate the response of hydrodynamics in San Francisco Bay for coastal protection infrastructure \cite{li2020efficient,jia2019investigation},  %In those studies, GPs used only 5\% of the simulation scenarios to approximate the behavior of a large-scale numerical model with an error less than 10\%. 
 and to forecast local wind fields in onshore wind farms \cite{ezzat2018spatio}. Here, we employed a GP with automatic relevance determination \cite{williams2006gaussian}, which naturally factors in the relative importance of the input variables into the regression model. Details about the mathematical foundations of GPs, and model training procedure, are deferred to Appendix B. \par

 %GPs have been used in engineering applications. For example, surrogates were trained to approximate the response of hydrodynamics in San Francisco Bay to coastal protection infrastructure using GPs\cite{li2020efficient,jia2019investigation}. In the studies, only 5\% of the simulation scenarios are needed to approximate the behavior of a large-scale numerical model with an error less than 10\%.
 
 %The stochastic simulation results are then analyzed by the Krigging method. Originated in geostatistics, Kriging or Gaussian Process regression is a method of interpolation assuming the interpolated values follow a Gaussian process governed by prior covariances. The method is widely used in spatial analysis and computer experiments. The technique is also known as Wiener–Kolmogorov prediction, after Norbert Wiener and Andrey Kolmogorov. Here, we use it as the major method to develop surrogate models and regression method to understand the sensitivity of the input parameters.

\section{Results}
We first present our model validation diagnostics of the OpenFAST simulations for the $15$ MW reference turbine, then proceed to conduct an exploratory analysis on the resulting stochastic simulations. An exhaustive sensitivity analysis is then performed to infer the impact of several environmental inputs on the offshore power and identify the most informative combination of inputs in light of the out-of-sample prediction accuracy. Finally, a case study is presented, wherein the trained models are evaluated using real-world offshore measurements from the New York/New Jersey Bight.

\subsection{\textbf{Validation of OpenFAST for the $15$ MW Reference Turbine Design}}

\begin{figure}[h]
    \captionsetup{justification=centering,margin=1cm}
    \centering
    \includegraphics[width=8.5cm]{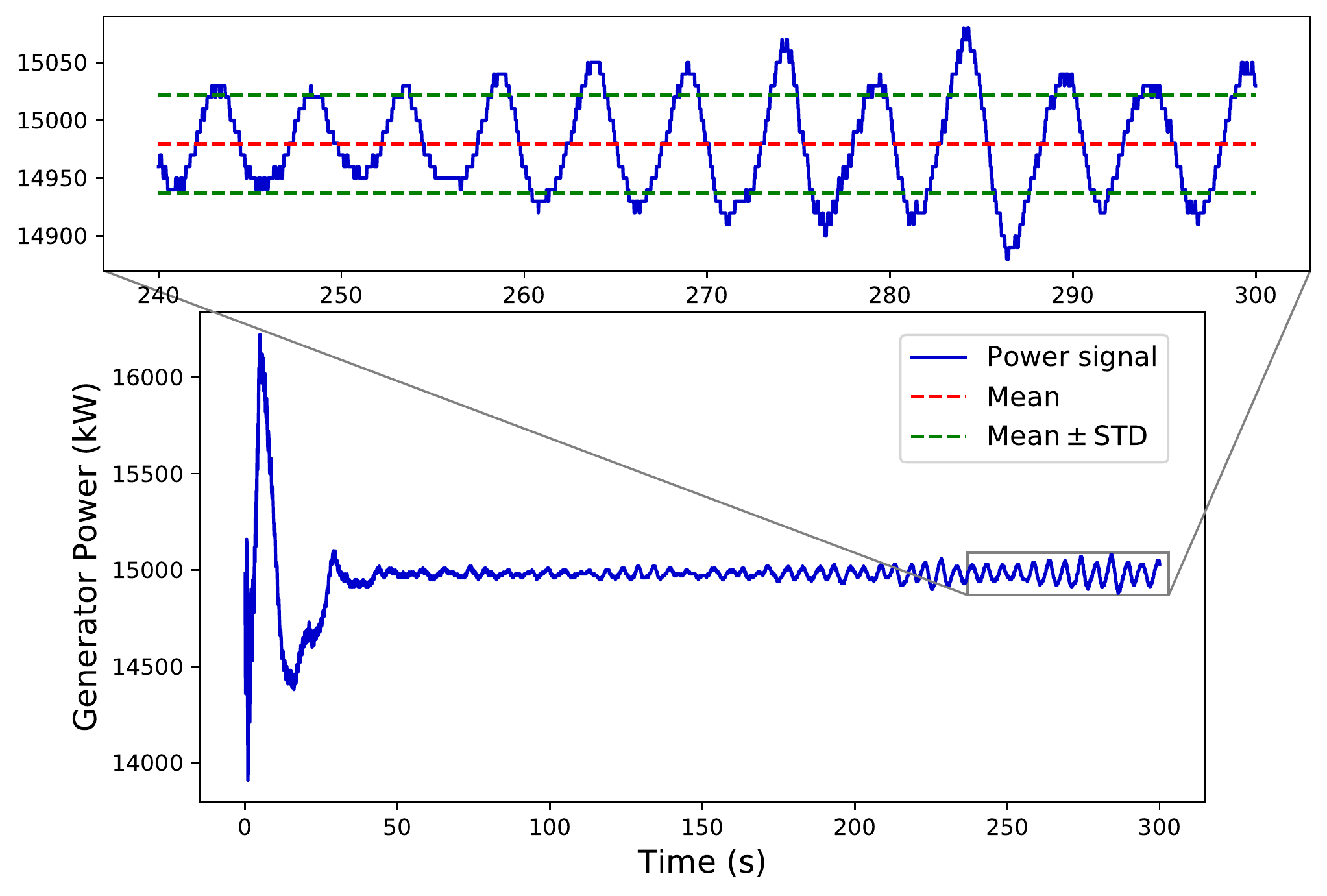}
    \caption{Generator power variations during a $300$-second simulation. As depicted, we only select the mean and standard deviation for the last $60$ seconds.}
    \label{fig:Mean&STD}
\end{figure}

\begin{figure}[!h]
    \captionsetup{justification=centering,margin=1cm}
    \centering
    \includegraphics[width=16cm]{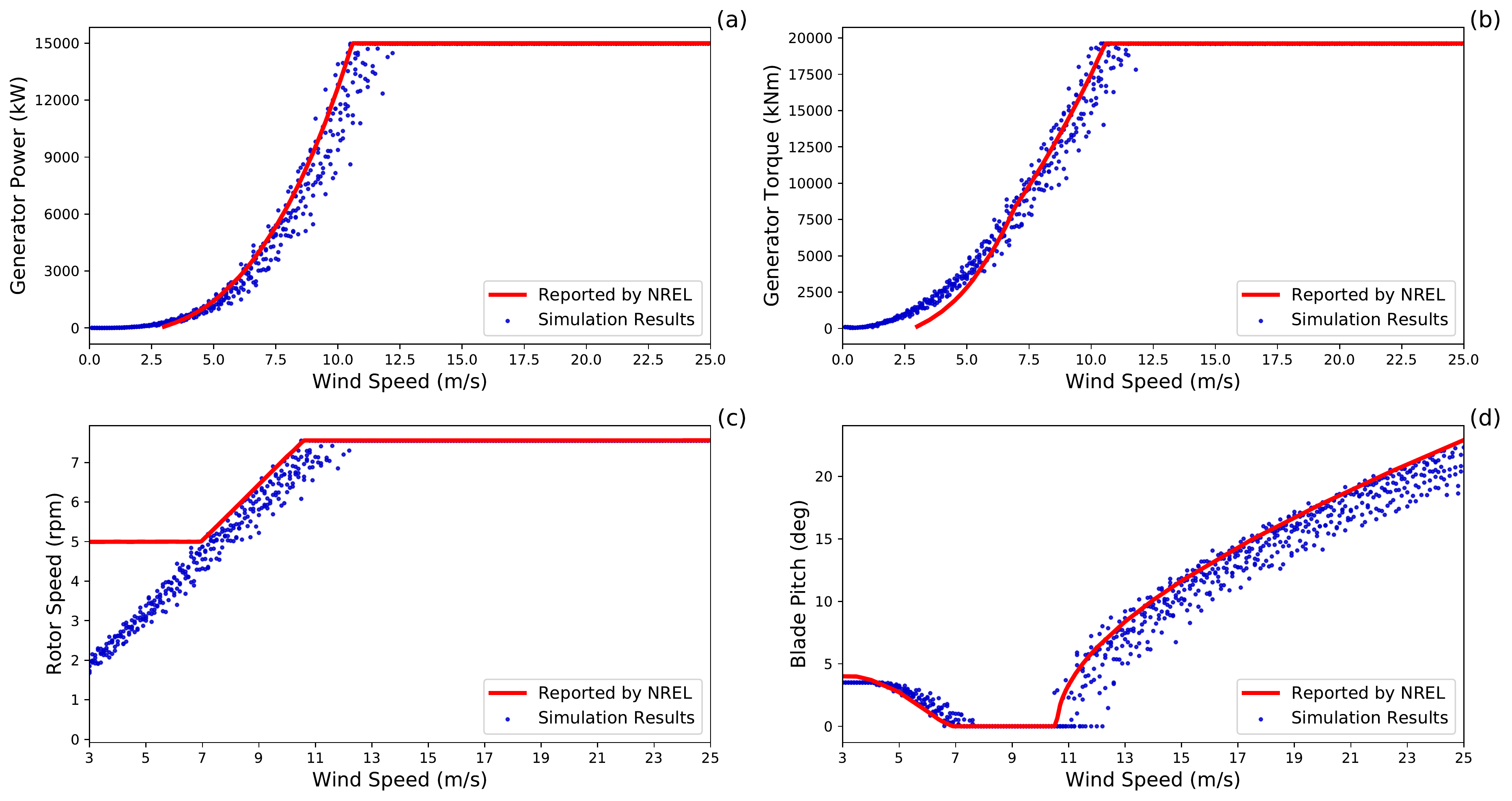}
    \caption{Comparison of steady-state values of the (a) generator power, (b) generator torque, (c) rotor speed, and (d) blade pitch angle with the data reported by NREL \cite{IEA15MW_ORWT}.}
    \label{fig:Validation}
\end{figure}

While OpenFAST has been validated in numerous studies for the $5$ MW NREL reference offshore wind turbine, its applicability to the new IEA $15$ MW turbine is not yet established. Here, we validate our OpenFAST simulation results using the nominal power curve for the  steady-state  performance of the turbine rotor presented in the NREL technical report \cite{IEA15MW_ORWT}. As the simulations in OpenFAST are transient, a total of $300$ seconds is modeled for each case and we only select the last $60$ seconds for analysis. The results have been visually examined for each case to ensure the spin-up time was not included in the data analysis. %The present study focuses on the time-averaging and standard deviation of the wind turbine behaviors, 
An example for the power output is illustrated in Fig. \ref{fig:Mean&STD}. %are considered for calculating the averaged value of each output parameter. This consideration has been taken into account to make sure unsteady effects caused by controller systems are eliminated. 
In Fig.~\ref{fig:Validation}, the time-averaging results are compared with the NREL technical data. As shown in Fig. \ref{fig:Validation}(a) and (b), the variations of generator power and torque with wind speed obtained from the stochastic simulations are in good agreement with the NREL data. In addition, the variations of rotor angular speed and blade pitch angle with the wind speed are plotted in Fig. \ref{fig:Validation}(c) and (d). 

It can be seen that for the rotor speed variations, there is a discrepancy between NREL data and steady-state results for wind speeds less than $7$ m/s. This difference can be attributed to the performance of the NREL Reference Open-Source Controller (ROSCO) \cite{ROSCO_2020} implemented in OpenFAST. In ROSCO, two proportional integral (PI) controllers for generator torque and blade pitch angle are actively running to ensure the wind turbine is operating within safe conditions under the design constraints, so the rotational speed of the turbine rotor is constrained to $5$ rpm and above \cite{IEA15MW_ORWT}. %It seems the PI controller of generator torque that is responsible for maintaining the minimum rotor speed contributes to the discrepancy in Fig. \ref{fig:Validation}. 

% \subsection{\textbf{Power curve modeling} and \textbf{Variations in power curve}}
%Fig. \ref{fig:Validation} also shows that the OpenFAST model didn't closely follow the nominal power curve when the wind direction is greater than 65 degree. The difference actually can be quite significant as shown in Fig. \ref{fig:power_curve_zoom}. Next, we focus on these cases to elucidate the dominant factors that determine the difference in the power output. 

%To investigate the reason for this deviation, we focus on the time series of the modeling results to reveal the mechanism behind this observation in the next section.
% \begin{figure}[htp]
%     \centering
%     \includegraphics[width=12cm]{power_curve2.pdf}
%     \caption{Power Curve}
%     \label{fig:power_curve}
% \end{figure}
\subsection{\textbf{Exploratory Analysis of Wind Power Variability in Complex Environments}}\label{sec:analytical}
\begin{figure}[!h]
    \centering
    \includegraphics[width=10cm]{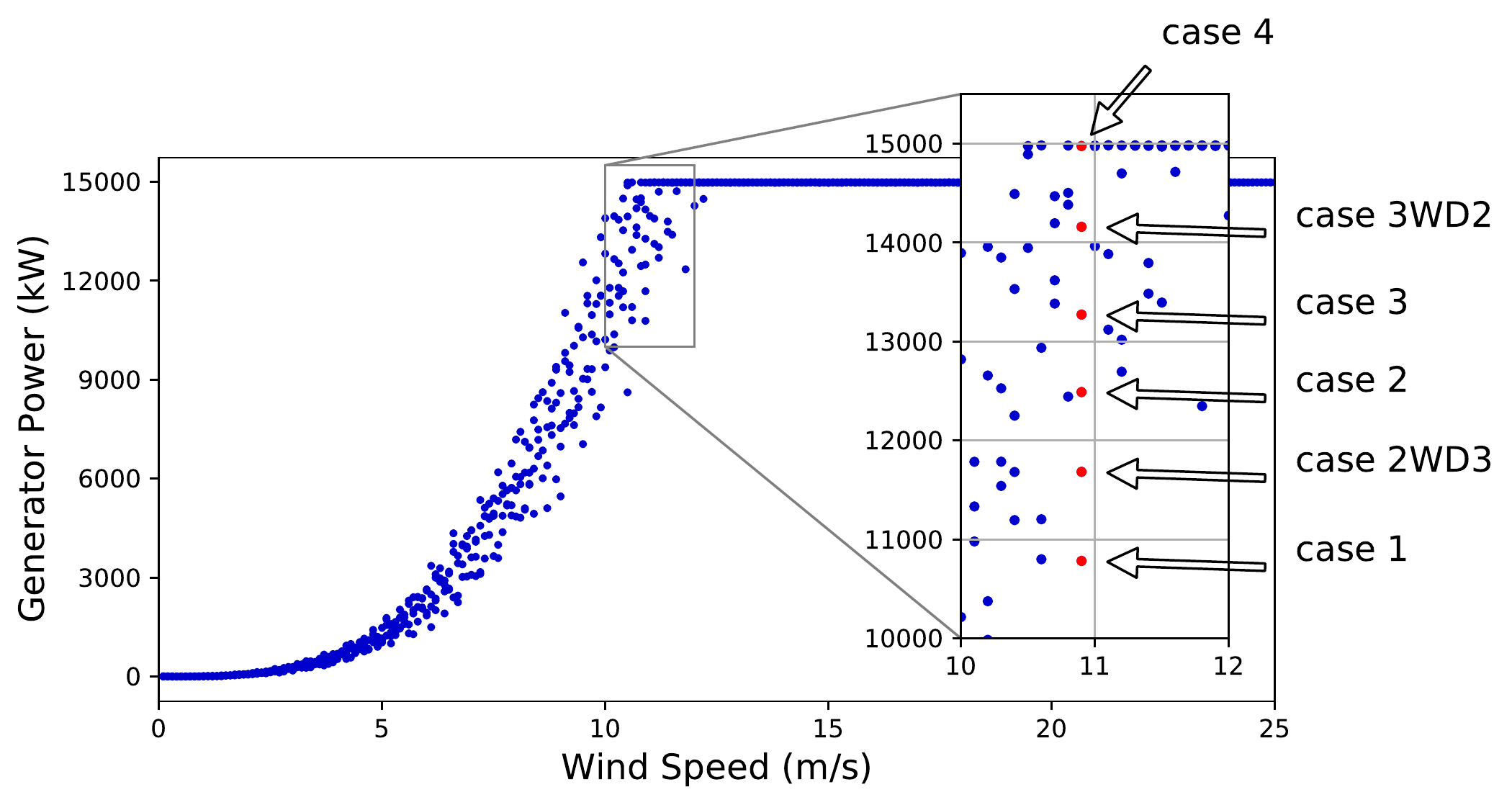}
    \caption{Results from OpenFast simulations depicting wind speed versus the resulting wind power. Inset highlights Cases $1$-$4$ which have similar wind speeds, but substantially different environmental conditions}
    \label{fig:power_curve_zoom}
\end{figure}

Wind speed is known to be the major determinant of offshore wind power, but other factors may exhibit a secondary, yet significant effect. This secondary effect is often overshadowed by the significant impact of wind speed variability on wind power. To examine the impact of other environmental factors, we specifically highlight four cases which have similar (almost identical) wind speeds of approximately $10.9$ m/s. Those are denoted hereinafter as Cases 1-4, and are depicted in Fig.~\ref{fig:power_curve_zoom}. %to eliminate the factor of wind speed for a comparison.
The environmental conditions for these cases are listed in Table \ref{tab:cases2}.  

\begin{table}[!h]
\centering
\caption{The environmental conditions for the four cases used for time series comparison}
\label{tab:cases2}
\begin{tabular}{>{\hspace{0pt}}m{0.08\linewidth}>{\centering\hspace{0pt}}m{0.06\linewidth}>{\centering\hspace{0pt}}m{0.09\linewidth}>{\centering\hspace{0pt}}m{0.080\linewidth}>{\centering\hspace{0pt}}m{0.1\linewidth}>{\centering\hspace{0pt}}m{0.08\linewidth}>{\centering\hspace{0pt}}m{0.12\linewidth}>{\centering\hspace{0pt}}m{0.09\linewidth}>{\centering\arraybackslash\hspace{0pt}}m{0.08\linewidth}} 
\hline
\multirow{3}{*}{Case} & Wind~  & Wind~      & Wave~   & Wave~      & Air~      & Air~         & Relative~ & Air~      \\
                      & Speed~ & Direction~ & Height~ & Direction~ & Pressure~ & Temperature~ & Humidity~ & Density~  \\
                      & (m/s)  & (deg)       & (m)     & (deg)       & (Pa)      & ($^\circ$C)          & (\%)      & (kg/m3)   \\ 
\hline
1                     & 10.9   & 28.3       & 2.4     & 111.4      & 97011   & 13.6        & 48.2      & 1.1756    \\
2                     & 10.9   & 16.9       & 1.3     & 78.8       & 97525   & 36.2        & 81.2      & 1.0778    \\
3                     & 10.9   & 20.8       & 5.2     & 129.6      & 111319  & 39.9        & 62.0      & 1.2200    \\
4                     & 10.9   & 5.8        & 13.3    & 70.1       & 104234  & 2.3         & 77.4      & 1.3162    \\
3WD2                  & 10.9   & 16.9       & 5.2     & 129.6      & 111319  & 39.9        & 62.0      & 1.2200    \\
2WD3                  & 10.9   & 20.8       & 1.3     & 78.8       & 97525   & 36.2        & 81.2      & 1.0778    \\
\hline
\end{tabular}
\end{table}

OpenFAST results for these cases show that the generated power ranges from $10.8$ to as high as $15.0$ MW, with the maximum being $\sim\hspace{-0.11cm}39\%$ higher than the minimum, suggesting that other environmental variables, in addition to wind speed, can have a significant impact on offshore power generation. A closer look reveals that two factors (other than wind speed) are dominant in determining the generator power: air density and relative wind direction. Specifically, by ranking the four cases according to their relative wind direction values, we can clearly see from Fig.~\ref{fig:power_curve_zoom} that Case 4 has the minimal relative wind direction value and the maximum wind power generation. This order is  reversed for Case 1, suggesting an inverse relationship between relative wind direction and generated power (contingent on a constant wind speed). Along the spectrum, Cases 2 and 3 have intermediate relative wind direction values and consequently, their correspondent wind power values are both smaller than that of Case 4, but larger than that of Case 1. We note, however, that Case 3 generates higher power than that of Case 2, albeit having a slightly greater relative wind direction value. We conjecture that this can be explained by the difference in air density, mainly driven by a large difference in air pressure. It is noted that the differences in other factors such as air temperature and relative humidity are less significant. To test this conjecture, we added two more cases: Case 3WD2 similar to Case 3, but with the relative wind direction of Case 2, and Case 2WD3 similar to Case 2, but with the relative wind direction of Case 3. 
It can be seen in Fig.~\ref{fig:power_curve_zoom} that Case 2WD3 generates less power than Case 2 that is obviously because of greater yaw misalignment; however, Case 3WD2 generates approximately 14\% higher power than Case 2 , suggesting that air pressure and temperature (and consequently, density) play key roles in describing offshore power generation even if wind speed and relative wind direction are similar in different conditions. This aligns with the physical knowledge from the power generation law $P = \frac{1}{2}C_p\rho A V^3$, where the air density depends on air pressure and temperature (See Eq. (\ref{eq:rho})), and the relative wind direction determines $V$. 

Although it is obvious that wind speed is the most significant factor in determining mean power, it is still unclear that whether yaw misalignment or air density is the second important factor under different environmental conditions. Considering the ratio of partial derivative of mean generator power with respect to yaw misalignment to the partial derivative with respect to density, we can examine the relative significance of these factors in describing mean power variability. As detailed in Appendix A (See Eq.~(\ref{eq:rat})), it can be shown that the absolute value of this ratio is $|(\partial P /\partial \theta)/(\partial P /\partial \rho) |=3\rho tan(\theta)$. Hence, the importance of the variable depends on whether the ratio is greater or less than 1. Given that widest possible ranges of values are considered in the simulations, if the maximum and minimum air density in our dataset are substituted for in the ratio's expression, two thresholds can be obtained for yaw misalignment as follows:
\begin{equation}\label{thresholds}
\begin{aligned}
    \theta_1 &=\arctan \left(\frac{1}{3\rho_{max}} \right) = 12.34 ~\textrm{deg} \\
    \theta_2 &=\arctan \left(\frac{1}{3\rho_{min}} \right) = 18.36 ~\textrm{deg},
\end{aligned}
\end{equation}
wherein if $\theta$ is less than $\theta_1$, it is guaranteed that the significance of density is higher than yaw misalignment, while if $\theta$ is greater than $\theta_2$, the generator power variability is more influenced by yaw misalignment rather than density. The cases with relative wind direction values between these two thresholds indicates the significance of density and yaw misalignment in predicting mean power variability are in the same order. %The subsets are illustrated in Fig. \ref{fig:thresholds}. %
It should be noted that in the whole dataset the number of cases with the ratio less than 1 is approximately the same as the number of cases with the ratio greater than 1. Therefore, using this analysis we cannot conclude which factor is the second dominant factor in describing wind power variability throughout the whole dataset. This is one of the limitations in conventional methods, thereby motivating the use of a data-driven approach particularly GP, which is able to infer the relative significance of input variables.      

\subsection{\textbf{Global Sensitivity Analysis using GPs}}
%While valuable in its own, using the analytical analysis of Section 3.2 to estimate the impact of all seven environmental variables on the generator power is difficult, %In contrast to the analytical analysis in the last section, it is difficult to estimate the impact of the individual environmental parameter on the generator power in our sensitivity study that involves all the seven environmental parameters, 
%because the power to input relationship is implicit and cannot be directly derived analytically. 
To collectively estimate the impact of all seven environmental variables on the generator power, we seek a statistical data-driven approach using GPs to statistically infer the sensitivity of the power output to the environmental factors. The idea here is to train multiple GP models on different combinations of the environmental variables, using the OpenFAST simulation data. The relative importance of the input variables is inferred by the correspondent change of the GP-based prediction errors with and without a particular variable. We adopt a $10$-fold cross validation scheme and use three error metrics to evaluate the prediction errors, namely: the Mean Absolute Error (MAE), Root Mean Square Error (RMSE), and Normalized Root Mean Squared Error (NRMSE). which are expressed in Eqs~(\ref{eq:mae})-(\ref{eq:nrmse}).
\begin{equation} \label{eq:mae}
\textrm{MAE}=\frac{1}{n} \displaystyle\sum_{i=1}^{n} |y_i-\hat{y}_i|,
\end{equation}
\iffalse
\begin{equation} \label{eq:mre}
\textrm{MRE}=\frac{1}{n} \displaystyle\sum_{i=1}^{n} \left[ \frac{|y_i-\hat{y}_i|}{y_i}\times 100 \right],
\end{equation}
\fi
\begin{equation}\label{eq:rmse}
\textrm{RMSE}=\sqrt{\frac{1}{n} \displaystyle\sum_{i=1}^{n} (y_i-\hat{y}_i)^2},
\end{equation}
\begin{equation} \label{eq:nrmse}
\textrm{NRMSE}=\frac{1}{\tilde{y}} \sqrt{\frac{1}{n} \displaystyle\sum_{i=1}^{n} (y_i-\hat{y}_i)^2},
\end{equation}
where $y$ and $\hat{y}$ denote the actual value and prediction of the predictand, respectively, while $\tilde{y}$ is a reference value, which will be specified depending on the response of interest (more details in Sections 3.3.1 and 3.3.2).

\begin{table}[b!]
\centering
\caption{Prediction errors for different multivariate power curve models in terms of averaged MAE, RMSE, and NRMSE in 10-fold cross validation}
\label{tab:newresults}
\begin{tabular}{llccc} \hline
No. &  Models                              & $MAE_{avg}$           & $RMSE_{avg}$          &  $NRMSE_{avg}$     \\
                                                                                                
  &                                                             & (kw)          & (kw)          & (\%)      \\\hline
1 &    Simple Binning                                           & 243.1         & 571.3         & 3.81      \\
2 &    Method of Bins (IEC Standard 61400–12-1)                 & 207.0         & 488.0         & 3.25      \\
3 &    $GP(V)$                                                  & 240.6	        & 529.2         & 3.53       \\
4 &    $GP(V, \theta)$                                          & 149.1         & 333.7         & 2.22       \\
5 &    $GP(V, \theta, T)$                                       & 120.5         & 257.6         & 1.72       \\
6 &    $GP(V, \theta, T, P)$                                    & 30.9          & 78.3          & 0.52       \\
7 &    $GP(V, \theta, T, P, \phi)$                              & 31.0          & 78.8          & 0.53       \\
8 &    $GP(V, \theta, T, P, \rho)$                              & 23.1          & 60.3          & 0.40       \\
9 &    $GP(V, \theta, T, P, \rho, H)$                           & 24.6          & 63.7          & 0.42       \\
10 &   $GP(V, \theta, T, P, \rho, H, \beta)$                    & 23.3          & 60.6          & 0.40       \\
11 &   $GP(V, \theta, \rho)$                                    & 23.1          & 60.3          & 0.40       \\\hline 
\end{tabular}
% \vspace{1ex}
% 
% {\raggedright * MEA=Mean Absolute Error, RMSE=Root Mean Square Error, MRE=Mean Relative Error \par}
\end{table}

\subsubsection{First-order properties: The sensitivity of the mean power output to the environmental variables}
To elucidate the effect of individual environmental inputs, a set of GP models is trained to the mean power output from the stochastic simulations, where, in addition to wind speed, we incrementally add the other environmental variables, one by one. We also include two benchmarks to compare against: simple binning and the standard method of bins. Simple binning takes wind speed as the sole input, and divides its domain into bins of $0.5$ m/s, which is the standard bin width in power curve estimation studies, and then takes the average of the power values within each bin as the estimated power. The method of bins, on the other hand, is similar to the binning method, but indirectly takes air density into account by performing a density correction to the wind speed values, as recommended by the IEC standard 61400-12-1 \cite{iec2005}.

\begin{figure}[!b]
    \captionsetup{justification=centering,margin=1cm}
    \centering
    \includegraphics[width=16.5cm]{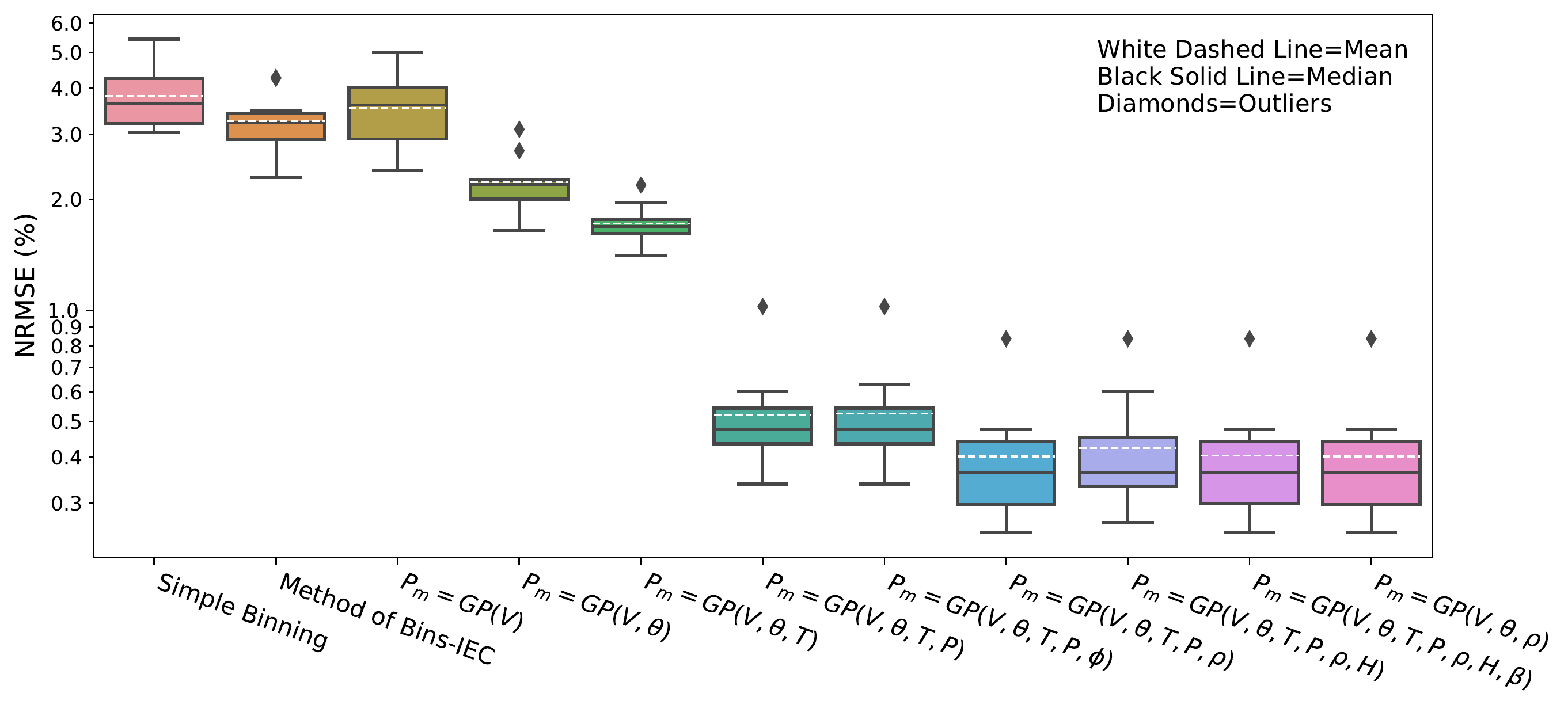}
    \caption{Boxplots of the NRMSEs of the mean power output predictions, across the 10 folds, for the $11$ different models thus considered.}
    \label{fig:boxplots1}
\end{figure}

Table \ref{tab:newresults} shows the final prediction errors of all models, across the three metrics of Eqs~(\ref{eq:mae})-(\ref{eq:nrmse}). For NRMSE, we use the rated power as the reference value, that is, $\tilde{y} = P_{rated} = 15,000$ kW. In general, as more input variables are added, prediction errors gradually decrease, except for the cases where adding relative humidity, wave height or wave direction, leads to a slight increase in prediction error. This is further illustrated in Fig. \ref{fig:boxplots1}, which depicts the boxplots of the NRMSEs, across the $10$ folds, for each model. Looking at Table \ref{tab:newresults} and Fig. \ref{fig:boxplots1}, we can infer the following key insights. 

First, it can be seen that the standard method of bins achieves a noticeable improvement of $14.6\%$ in comparison with simple binning. It is also shown that, by leveraging the local correlation structure in the input space, a GP model with wind speed as the sole input, denoted by $GP(V)$, outperforms the simple binning method ($\sim\hspace{-1mm}7.4\%$ improvement in NRMSE), but still falls short to the standard method of bins, which indirectly augments the input space via its embedded density correction. This finding indicates that, even advanced statistical approaches like GPs may not be able to provide significant competitive advantages over traditional approaches when solely using the wind speed as the only input\textemdash a conclusion which is largely overlooked in the literature, barring few recent works \citep{lee2015power,pandit2019incorporating}. This aligns with our findings from the exploratory analysis in Section 3.2, and highlights the importance of considering other environmental variables in addition to wind speed in characterizing offshore wind power.

Second, by using a GP model with wind speed and relative wind direction as input variables, a significant improvement ($\sim\hspace{-1mm}31.7\%$) in comparison with the method of bins is observed, indicating that relative wind direction plays a key role in describing the mean power output, which again, coincides with our exploratory analysis in Section 3.2. By training a four-variable GP model and considering air temperature and pressure, $GP(V, \theta, T, P)$, significantly higher accurate predictions are further achieved ($76\%$ improvement in NRMSE relative to the method of bins). The best five-variable model is $GP(V, \theta, T, P, \rho)$, which achieves a final NRMSE of $0.4$, which is $23\%$ better than its four-input predecessor, and up to $88\%$ better than the method of bins. Although density is highly correlated to air pressure and temperature, we speculate that the realized improvement may be attributed to the fact that density can further capture the interaction effects of temperature and pressure. We alo note that adding wave height and wave direction does not lead to any significant improvements in predictive performance, concluding that the wind-related variables are the dominating environmental factors in describing the mean power output, while, on the other hand, the wave properties do not have a significant contribution, which is, perhaps, unsurprising.

Third, as thoroughly described in Appendix B, we employed an ARD-SE covariance function (See Eq.~\ref{eq:gp2}), in which the estimated values for length-scale parameter $(\ell)$ can reveal the importance of each input variable, such that smaller value (or higher inverse values) of $\ell$ suggests higher significance. By normalizing 1/$\ell$ values, the contribution of each variable in three select GP models are listed in Table \ref{tab:1overL}. For all three models, wind speed and yaw misalignment have the first and second highest importance with contributions of about $92\%$ and $5\%$, respectively. Furthermore, it can be inferred that in the $GP(V, \theta, T, P)$ model the importance of air pressure is slightly higher than air temperature. Also, it can be seen that, once density ins included in $GP(V, \theta, T, P, \rho)$ model, its contribution is about $3\%$, while the contributions of air temperature and pressure drop to around zero. This means that density appears to effectively encapsulate the combined impact of air pressure and temperature, and indicates that the best model is $GP(V, \theta, \rho)$, which (i) achieves the best performance in terms of prediction error across all metrics ($\sim\hspace{-1mm}88\%$ reduction in NRMSE relative to the method of bins), (ii) is well aligned with the physical law of power generation, and (iii) is further in line with the law of parsimony in statistical and machine learning, particularly when compared against other model representations with more variables (and hence higher complexity), but offer no competitive advantage in terms of prediction accuracy.

\vspace{6mm} %5mm vertical space

\begin{table}[!h]
\centering
\caption{Inverse of length-scale parameter $(\ell)$ and the percentage of the importance of input variables in GP regression}
\label{tab:1overL}
\begin{tabular}{lcccccccc}
\hline
                 & \multicolumn{2}{c}{$GP(V, \theta, T, P)$} &
                 & \multicolumn{2}{c}{$GP(V, \theta, T, P, \rho)$} &
                 & \multicolumn{2}{c}{$GP(V, \theta, \rho)$} \\ \cline{2-3} \cline{5-6} \cline{8-9} 
                 Input Variable & 1/$\ell$                  & \%                 &
                 & 1/$\ell$                      & \%                   &
                 & 1/$\ell$                 & \%                  \\ \cline{1-6} \cline{8-9} 
Wind Speed $(V)$    & 5.921    & 92.6    &  & 6.656    & 91.92    &  & 6.656    & 91.92               \\
Yaw Misalignment $(\theta)$    & 0.309    & 4.84    &  & 0.368    & 5.09     &  & 0.368    & 5.09                \\
Air Pressure $(P)$    & 0.091    & 1.43    &  & \num{1e-6}    & $\approx0$    &  & -        & -                   \\
Air Temperature $(T)$    & 0.067    & 1.06    &  & \num{5e-6}    & $\approx0$    &  & -        & -                   \\
Air Density $(\rho)$    & -        & -       &  & 0.216    & 2.99     &  & 0.216    & 2.99                \\ \hline
\end{tabular}
\end{table}

\vspace{3mm} %5mm vertical space

\subsubsection{Second-order properties: The sensitivity of the variance in power output to the environmental variables} \label{powerSTDmodelprediction}
The second order properties of the power output, defined in this paper as the variation of the power around its mean, is an important, but often overlooked, quantity in power curve estimation studies. As a case in point, the high variance in power output can cause significant disruptions to grid integration and energy storage. Moreover, it contributes to structural fatigue, foundation instability, and marine noise pollution. %The IEC Standard 61400–12-2 suggests to use the empirical value of 2\% of the power output as the magnitude of uncertainty caused by changing climatic conditions. 
The IEC Standard 61400–12-2 suggests that the uncertainty in the power curve may be on the order of $10$\% or more, depending on site conditions and climate \cite{iec2013}. Understandably, the uncertainty may be affected by a combination of several environmental factors. Here, we seek to infer the impact of these factors on the variation of the power output around its mean behavior by analyzing the results of the stochastic simulations generated by OpenFAST. 

\begin{figure}[h]
    \captionsetup{justification=centering,margin=1cm}
    \centering
    \includegraphics[width=16.5cm]{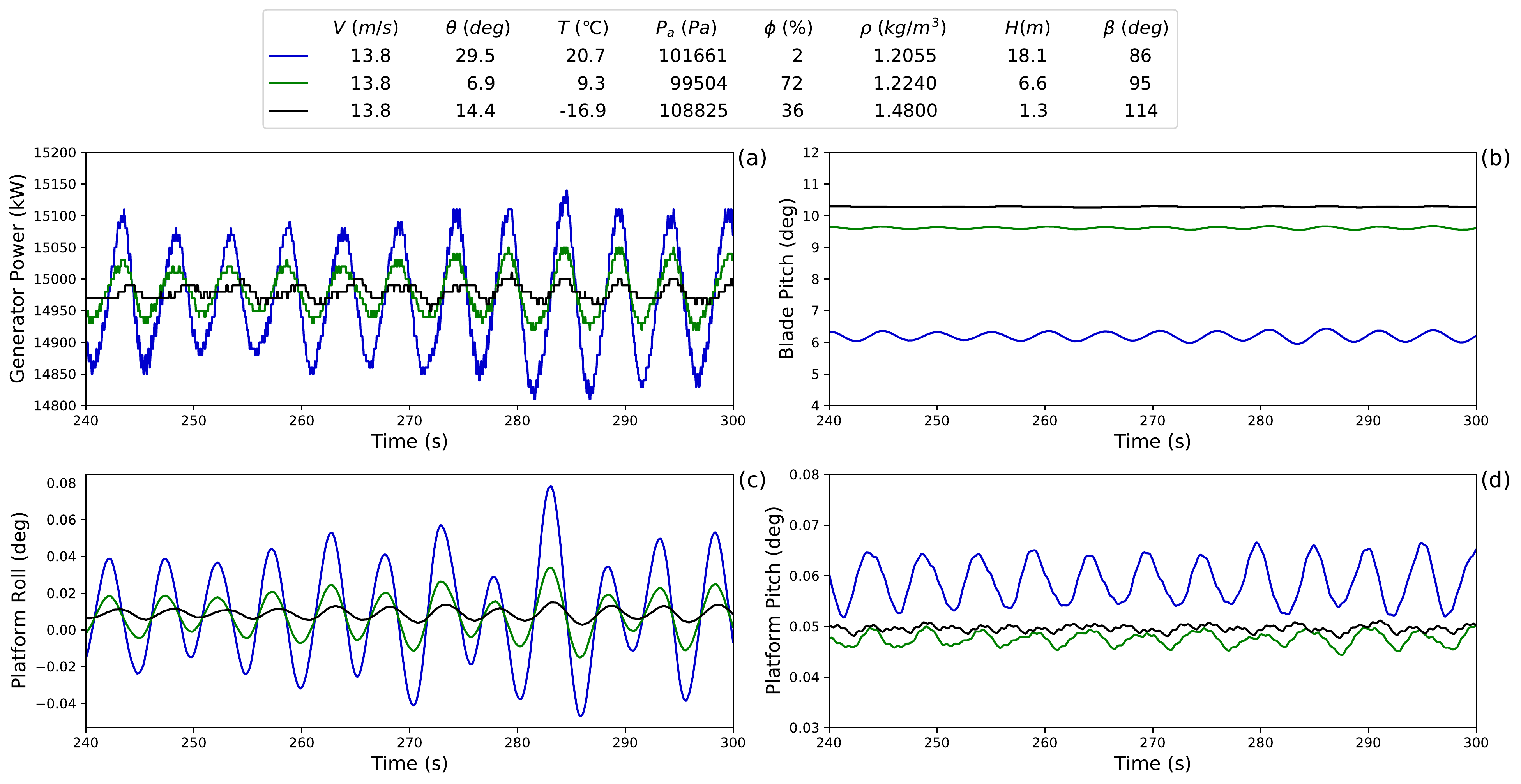}
    \caption{Fluctuations of (a) output power, and the variations of turbine (b) blade pitch, (c) platform roll, and (d) platform pitch during one-minute for three cases with identical wind speed and mean output power}
    \label{fig:TimeSeriesResults_NEW}
\end{figure}

To motivate the analysis, we particularly highlight three simulated cases that have identical wind speeds, and approximately the same mean power output, yet they exhibit extremely different behaviors in terms of the variability around that mean. Fig. \ref{fig:TimeSeriesResults_NEW} shows sample results for the time variations in the power output and the three displacements of the turbine platform, namely roll, pitch, and blade pitch angle for these three cases. The definition of the rotational displacements are depicted in Fig. \ref{fig:Schematic}. The table in the top panel of Fig. \ref{fig:TimeSeriesResults_NEW} shows the values of the relevant environmental variables for these three cases. As can be seen in Fig. \ref{fig:TimeSeriesResults_NEW}, the case in which the turbine experiences stronger displacement fluctuations has larger periodic variations around the mean power output. Interestingly, we observe that the amplitude of these power variations may be positively correlated with wave height, $H$, that is, larger wave heights correspond to more severe power variations.

\begin{figure}[h]
    \captionsetup{justification=centering,margin=1cm}
    \centering
    \includegraphics[width=15.25cm]{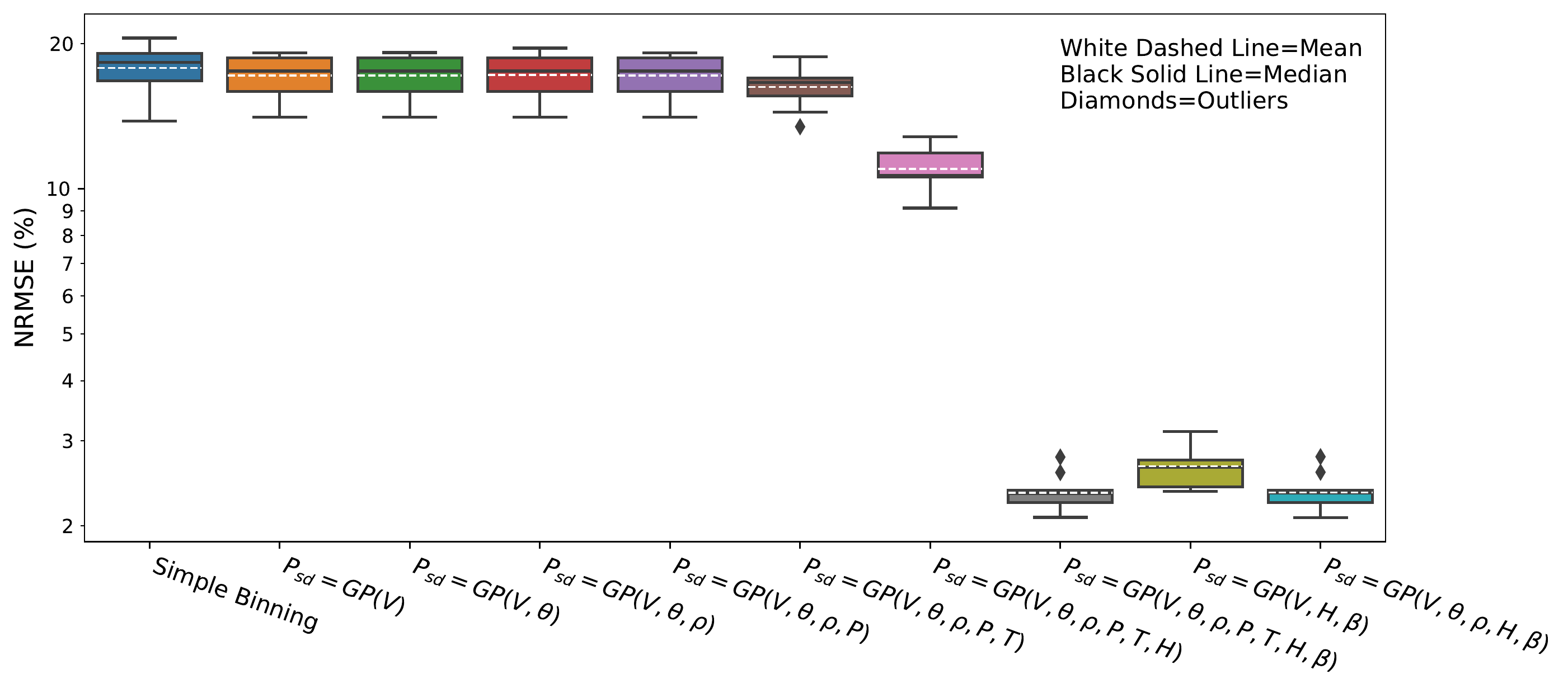}
    \caption{Boxplots for the distributions of normalized root mean square errors in 10-fold cross validation training and predicting generator power standard deviation for different multivariate GP models and simple binning method.}
    \label{fig:boxplots2}
\end{figure}

To test this conjecture, we perform a similar sensitivity analysis to that of Section 3.3.1. to study the second-order properties of power output. Here, we use the standard deviation of the generator power, denoted by $P^{sd}$, to quantify the second moment and the reference value used in NRMSE calculations is the maximum observed standard deviation in the training data, i.e. $\tilde{y} = \underset{i=1,...,n}{\max} P^{sd}_{i}$. The results, shown in Fig.~\ref{fig:boxplots2}, indicate that wind-related variables, alone, cannot explain the variance in the power output. Once the wave-related variables are added as inputs, the predictive performance is significantly improved. In particular, the model with wind speed, wave height, and wave direction, denoted by $GP(V, H, \beta)$, achieves $\sim\hspace{-1mm}86\%$ improvement, in terms of NRMSE, over the model which uses wind speed as the sole input, namely $GP(V)$. Minor to null improvements are achieved by adding additional variables beyond those three variables, which, again, is well aligned with the law of parsimony. This analysis interestingly suggests that wave-related variables are significant contributors to the power output uncertainty in offshore environments, whether as individual effects, or as interaction with the wind speed variable. At the first glance, the strong importance of the wave effects may be a little surprising since they are insignificant in characterizing the mean power output, as shown in Section 3.3.1. This finding can be explained by the fact that periodic waves are the only dynamic forcing among all the environmental variables under consideration. 

In addition to $GP(V, H, \beta)$ which demonstrates a satisfactory performance in predicting the second order properties of the power output, %but the model is implicit in mathematical expression -- the users have to apply the model to learn the importance of each input parameter. In addition, the model is transferred in the format of a dataset/computational code, which prevents a quick rule-of-the-thumb estimate. 
we develop a second-order effect curve, a counterpart to the ``power curve''  in Fig.~\ref{fig:power_curve_zoom}, which is based on fractional polynomial regression equation, shown in Eq.(\ref{eq:pol}), for which the parameters are estimated using non linear least squares. 
% \begin{equation}
%   \left[\frac{\sigma_P}{\frac{1}{2}{\rho}A{V_{cut-out}^3}}\right]=a+b{\mathit{(WSpd)}}^c+d{\mathit{(WvH)}}^e+f{\mathit{(WvDir)}}^g
% \end{equation}
\begin{equation}
%  \left[\frac{P_{sd}}{\frac{1}{2}{\rho}A{V_{cut-out}^3}}\right]=-3.63+2.70{\mathit{V}}^{0.38}+1.82{\mathit{H}}^{1.21}+{0.89}{\mathit{\beta}}^{0.01},
  \left[\frac{P^{sd}}{\frac{1}{2}{\rho}A{V_{cutout}^3}}\right]=-2.63+2.78{\mathit{\left(\frac{V}{V_{cutout}}\right)}}^{0.37}+1.81{\left(\frac{\mathit{H}}{\sqrt{A}}\right)}^{1.21}-{1.02}{\mathit{\beta}}^{4.86},
  \label{eq:pol}
\end{equation}
where $P^{sd}$ is the standard deviation of the generator power. The corresponding $10$-fold NRMSE of this model is $13.44\%$, which is still $\sim\hspace{-1mm}22\%$ better than that of $GP(V)$, albeit substantially higher than $GP(V, H, \beta)$. The main reason of introducing this model is its interpretability and ease of use, which may counterbalance its lower accuracy relative to $GP(V, H, \beta)$. From the fitted equation, we can note that the three factors have notable impact on $P^{sd}$, as evident by the estimated parameter values. 

Note that the present study focuses on the seven selected environmental factors, without including the effects of wind turbulence, shearing, and vaning, etc, so the strong wave effect may be less important than those dynamic forces in the offshore environment, which is indeed a topic of future research. We also stress that the present study is restricted to the dynamics that have been implemented in the OpenFAST framework, so any observation here is limited by the assumptions made in the OpenFAST framework. In turn, the developed GPs have the same limitations as OpenFAST, because the GPs are trained based on the OpenFAST results. But the limitations of the GPs are not permanent due to the independence between OpenFAST and GPs -- we expect that improved simulations or rich field observation data can reduce the limitation of the GPs. The analysis above reveals the potentially significant impact of ocean waves on the steadiness of the power output. The magnitude of the absolute fluctuation observed here can be compared with future studies focusing on the fluctuation introduced by wind turbulence, shearing, and vaning to identify the most influencing set of environmental variables. To the knowledge of the authors, this is the first attempt in a scientific study to particularly analyze, in depth, the determinants of the uncertainty in offshore power output, which can be potentially instrumental to grid integration and energy storage decisions. %ii) the study particularly highlights the potential advantage of OpenFAST than the power curve and simple analytical approaches in obtaining the fluctuation characteristics, and iii) the same method can be directly applied to the vibration of turbine structural components, which will dictate the material fatigue and life span. Nevertheless, 

%In practice, this equation can be directly applied to estimate the power output uncertainty to inform the operator about grid integration, storage, and structure stability.

% \begin{figure}[h!]
%     \captionsetup{justification=centering,margin=1cm}
%     \centering
%     \includegraphics[width=8.5cm]{PowerSTDvsVariablesForLinearReg.pdf}
%     \caption{Generator power standard deviation distribution plotted versus (a) wave direction, (b) wave height, and (c) wind speed obtained from MCS results}
%     \label{fig:Yaw}
% \end{figure}

%\hl{[Comments by Aziz: (1) is there a particular justification for the choice of the parametric form in Eq 8? (2) I assume S, A, B, C, and D are model parameters, what were their values, and how were they estimated? Can they provide any additional physical insights that we can discuss? (3) Have we tried to use Gaussian processes as well here? If yes, what was the outcome?]}

\subsection{\textbf{Case Study: Application to Real-world Data from the New York/New Jersey Bight}}

\begin{figure}[h!]
    \captionsetup{justification=centering,margin=1cm}
    \centering
    \includegraphics[width=16.5cm]{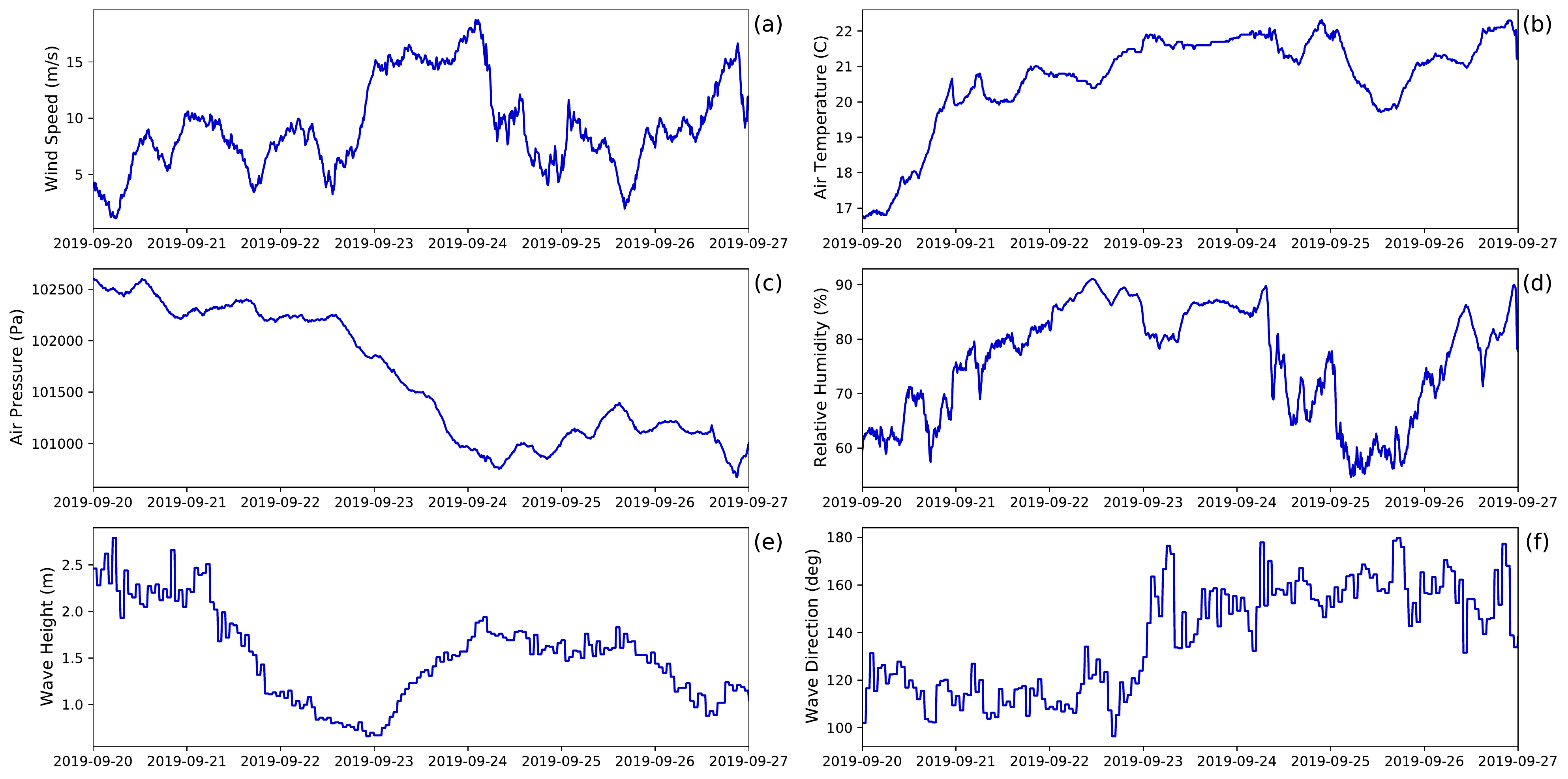}
    \caption{Variations of environmental variables during the $7$-day period recorded from NYSERDA's buoy dataset, including: a) wind speed, b) air temperature, c) air pressure, d) relative humidity, e) wave height, and d) wave direction.}
    \label{fig:envVariations}
\end{figure}

In this section, the best-performing models from Sections 3.3.1 and 3.3.2 for the prediction of mean and standard deviation of generator power, respectively, are applied to a real-world dataset collected using the E06 Hudson South LiDAR buoy, which is one of the buoys deployed by NYSERDA (New York State Energy Research \& Development Authority), strategically located in proximity to at least 3 future offshore wind farm locations, a few miles off the New Jersey and New York shorelines ($39^o 32' 48.38"$N, $73^o 25' 44.01"$W). Wind and wave data are collected, at multiple altitudes, since September 2019, and are made publicly available by NYSERDA \cite{nyserda2019}. Here, we focus on a 7-day period from September 20th to 27th, 2019, which includes $10$-minute measurements of the wind speed at the height of $158$ m (nearest height to the turbine hub height), air temperature, atmospheric pressure, and relative humidity. In addition, $1$-hour measurements of the wave height and wave direction are available during the same period. As shown in Fig.~\ref{fig:envVariations}(a), the wind speed during this period has a clear daily cycle. The weather experienced an early temperature and humidity rise, accompanied by a low pressure system in the later part of the 7-day period. Wave heights were mostly moderate with a visible shift in wave direction taking place approximately during the fourth day. 

\begin{figure}[h]
    \captionsetup{justification=centering,margin=1cm}
    \centering
    \includegraphics[width=9cm]{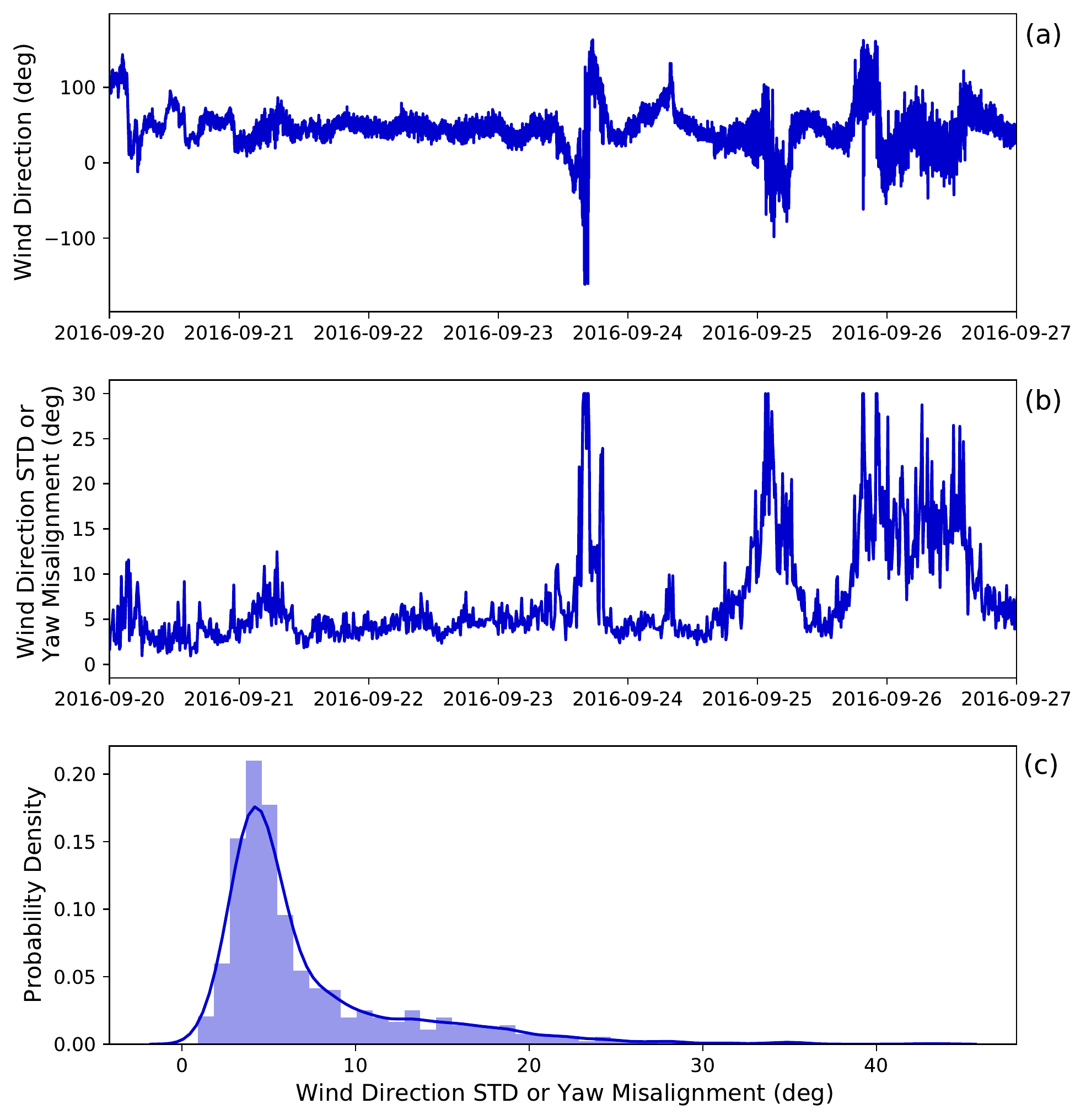}
    \caption{Analysis of the $1$-Hz AXYS Wind Sentinel Buoy data: (a) Time series of the wind direction, b) $10$-min wind direction standard deviation. c) Probability density of the wind direction standard deviation.}
    \label{fig:Yaw}
\end{figure}

The input data for OpenFAST can be determined using the buoy's field data, except for one variable: the relative wind direction (or yaw misalignment), since no wind farms exist yet at this location. Gaumond \textit{et al.} \cite{doi:10.1002/we.1625} suggested that the yaw misalignment is highly correlated with the second-order behavior of wind direction. This is due to the design of yaw controllers in modern-day horizontal wind turbines in avoiding the instantaneous tracking of incoming wind directions. In other words, if the wind direction varies more frequently and with higher amplitude, a wind turbine will experience greater yaw misalignment degrees. 
Hence, to estimate the yaw misalignment, we need to assess the standard deviation  \textit{within} $10$-minute periods (The $10$-min sampling frequency of the NYSERDA data is too coarse). %Following their method, argument, we estimate the relative wind direction using the wind magnitude and direction data. %A challenge in this task is the requirement of the wind sampling rate.%According to the IEC standard for large wind turbines \cite{doi:10.1002/we.1579}, 10-min period is recommended to be used to average the wind data for operation purposes. %it is not appropriate to produce the wind spectrum data. %, and there is report that the yaw misalignment for each 10-minute intervals can be a function of wind direction standard deviation within that 10-minute period. It was also indicated that the histogram of wind direction variations with the sampling rate of 12 Hz within an averaging time can be fitted by a normal distribution curve. %Therefore, for estimating the yaw misalignment values for the case study another dataset with higher measuring resolution is needed. Due to the limitation of available data 
To address this issue, we used an additional database -- the 1-Hz dataset collected by AXYS Wind Sentinel Buoy Lidar \cite{osti_1406993, osti_1632348}, which was deployed by Pacific Northwest National Laboratory (PNNL) from November 2015 to February 2017 in the New Jersey offshore area about 3 miles off the coast of Atlantic City, relatively near to the NYSERDA buoy location. %Because this dataset does not have wave data and its wind data had not been measured at the turbine hub-height, 
Specifically, we used this dataset for analyzing wind direction variations in the same 7-day period in 2016, shown in Fig. \ref{fig:Yaw}(a). The standard deviation of each 10-minute interval is calculated and plotted in Fig. \ref{fig:Yaw}(b), which will be used as a proxy for relative wind direction (or yaw misalignment). % in which only a few of them are higher than 30 degrees that is greater than the maximum yaw misalignment for which the models were trained on. For this reason, these out-of-range values are clipped to the maximum to make the data ready to be used for prediction.
The probability density of the yaw misalignment values is shown in Fig. \ref{fig:Yaw}(c), where most values are in the range of 0-30 degree.

\subsubsection{Case Study Results}

\begin{figure}[h!]
    \captionsetup{justification=centering,margin=1cm}
    \centering
    \includegraphics[width=16.5cm]{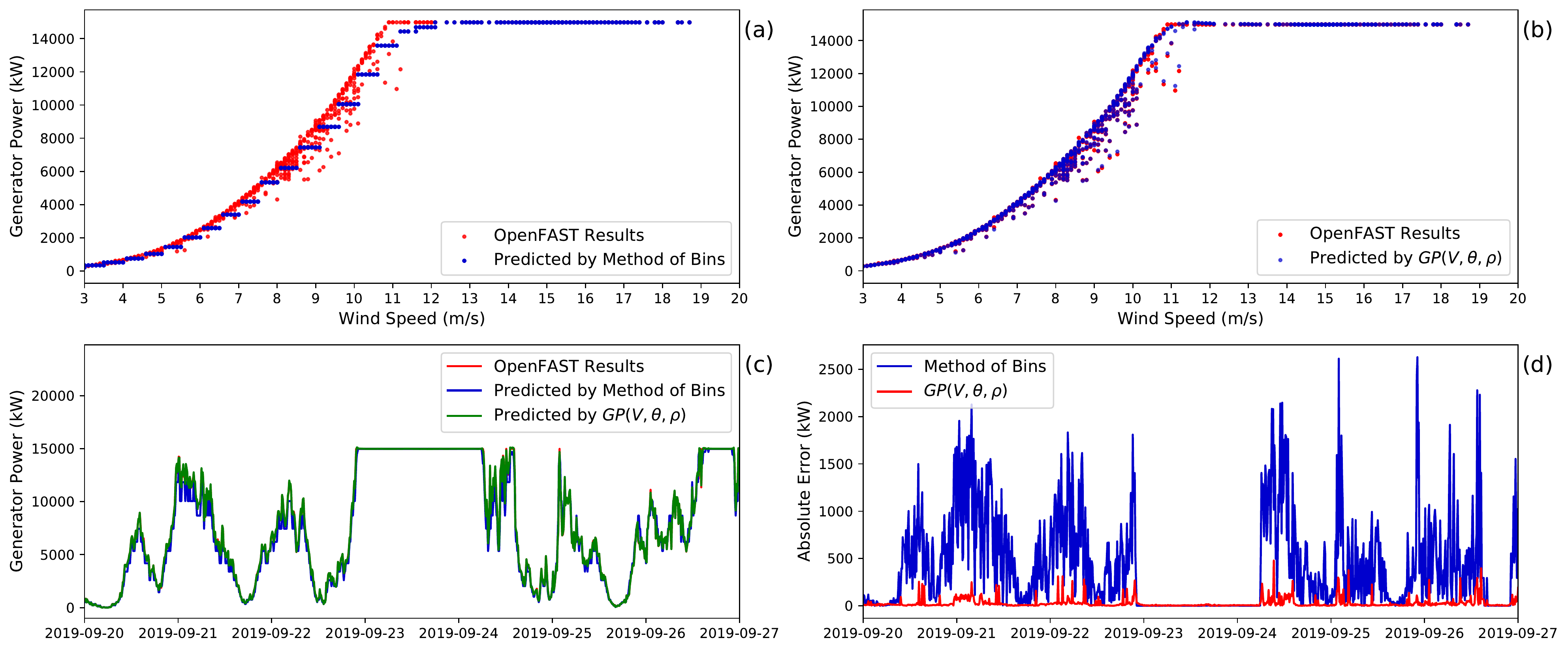}
    \caption{Comparison between OpenFAST simulation results and generator power values predicted by (a) method of bins, and (b) Gaussian Processes model of $GP(V, \theta, \rho)$; (c) Time series of generator power simulated by OpenFAST versus those predicted by the method of bins and $GP(V, \theta, \rho)$; (d) Time series of the absolute errors from $GP(V, \theta, \rho)$ versus those from the method of bins.}
    \label{fig:PredPower}
\end{figure}

OpenFAST is used to simulate the power generation for the buoy's location during the $7$-day period of interest. The GP models are then trained to the resulting simulation outputs. Fig. \ref{fig:PredPower}(a) and (b) show the power curves obtained by the method of bins and the best-performing GP model, respectively. % wherein we can visually note that the former is significantly less accurate, particularly for the middle region of the power curve, i.e., at wind speeds less than the turbine rated speed ($10.6$ m/s) and greater than $8.0$ m/s. However, the performance of this model is good enough for wind speeds higher than the turbine rated speed. On the other hand, 
 Fig. \ref{fig:PredPower}(b) shows that the generator power values predicted by the best three-variable GP model, namely $GP(V, \theta, \rho)$ are in excellent agreement with those obtained by OpenFAST, meaningfully better than the method-of-bins model, especially in the middle region of the power curve, i.e., at wind speeds less than the turbine rated speed ($10.6$ m/s) and greater than $8.0$ m/s. 

Fig.~\ref{fig:PredPower}(c) shows the time series of the generator power values simulated via OpenFAST, versus those predicted from the method-of-bins and $GP(V, \theta, \rho)$. The times series of the absolute errors of the two models are shown in Fig. \ref{fig:PredPower}(d), in which $GP(V, \theta, \rho)$ is shown to be $91\%$ more accurate than the method of bins, in terms of NRMSE. In terms of energy production, the total output energy predicted by OpenFAST over the one-week period is about 1347 MWh, while the total absolute energy error of the GP and method-of-bins relative to the OpenFAST total output energy are $4.6$ and $68.0$ MWh, respectively. Note that the selected period is not necessarily exhaustive nor representative of the life cycle of the wind turbine, and hence, we expect that the total energy prediction error may change in other weeks, seasons, or years. %when more moderate wind speed events are present, while the MRE is expected more reliable to scale up. The 6.1\% difference in MREs between the GP and method-of-bins models can be significant in the industrial application. 
The improvement in predictive performance achieved by the multi-input GP over the method of bins can be significant in industrial applications, and hence, we conclude that considering the environmental variables in addition to the wind speed, particularly the relative wind direction and air density, can significantly improve the prediction of offshore wind power.\par  

\begin{figure}[t!]
    \captionsetup{justification=centering,margin=1cm}
    \centering
    \includegraphics[width=9cm]{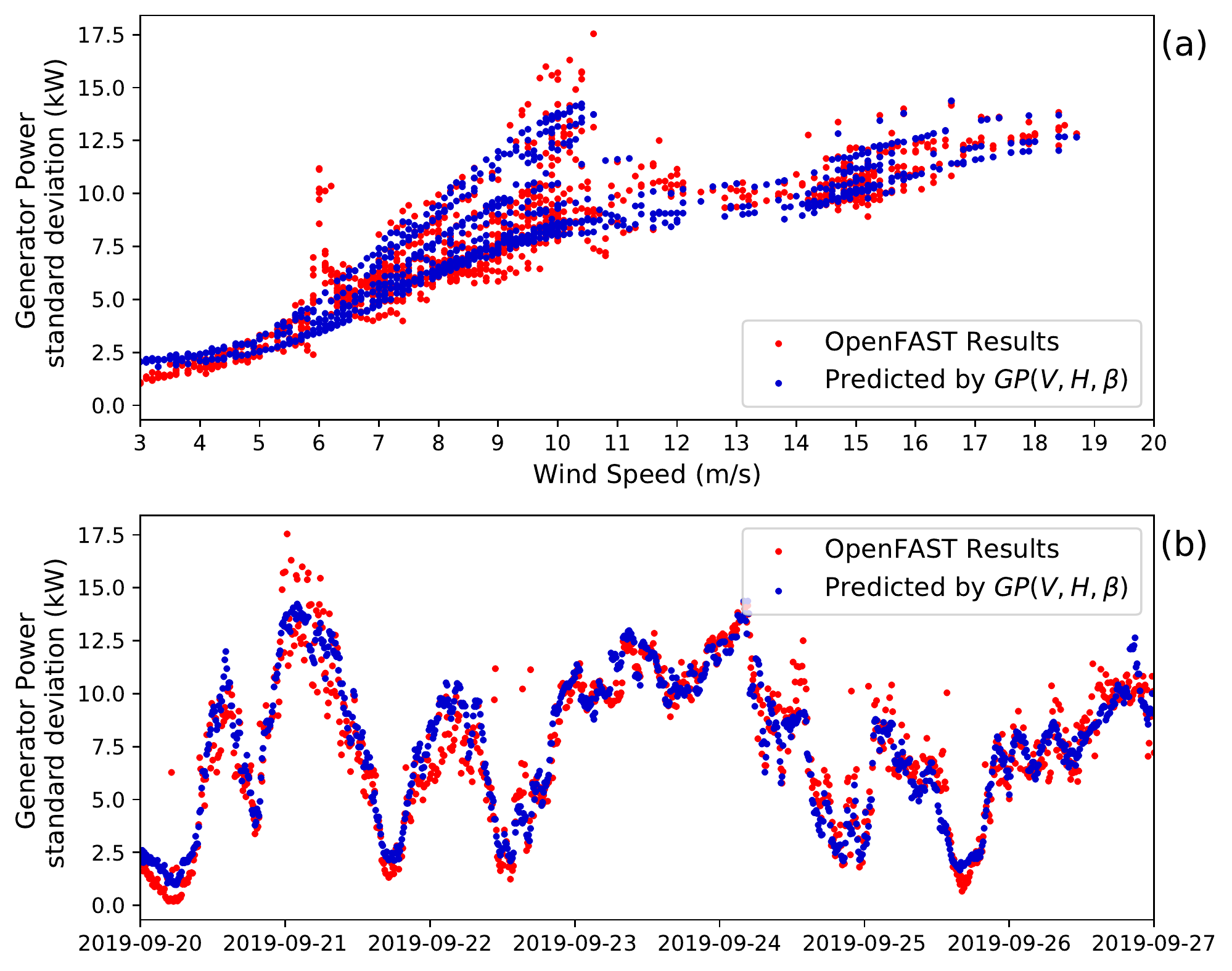}
    \caption{Comparison between the OpenFAST results and GP-based predictions for (a) different wind speeds, and (b) versus time.}
    \label{fig:PredPowerSTD}
\end{figure}

In addition, the standard deviation of the generator power is predicted for the same period using $GP(V, H, \beta)$. The power uncertainty counterpart of the power curve for the $7$-day period is shown in Fig.~\ref{fig:PredPowerSTD}(a). The GP-based predictions show a decent agreement with OpenFAST simulations, albeit not as good as those shown for the mean power prediction, which is expected. Particularly, we note that the prediction errors are higher at wind speeds in the range of $6$ to $10$ m/s. 
In Fig.~\ref{fig:PredPowerSTD}(b), the time series of the generator power standard deviation simulated by OpenFAST are shown against those predicted by the multi-input GP model. By comparing this Figure with Fig.~\ref{fig:PredPower}(c), it can be seen that during the time when the turbine generates the rated power, the predicted values of power standard deviation have the lowest errors, which, again, is unsurprising. 
% Overall, the MAE of the GP model in predicting power standard deviation is about 0.9 kW during the period, which is even lower than that obtained in cross-validation training and testing on the sensivitity analysis (see Fig. \ref{fig:boxplots2}): the MRE of the model on the case study dataset is about 13.1\%.

The case study shows that the statistical models proposed herein are well-suited to accurately predict the first two moments of the offshore power output. It also suggests that carefully integrating more environmental variables can result in substantial improvements to the prediction accuracy. %We also developed the first statistical and empirical models that are able to estimate the power generation uncertainty due to environmental variations. 
We envision our analysis to be particularly relevant to the growing offshore wind industry since an accurate prediction of the mean power output is known to be instrumental for productivity assessment, and for turbine- and farm-level operational activities. On the other hand, the capability to accurately assess the variation of the offshore power output around its mean, i.e. its second-order properties, can prove extremely useful to critical farm-level decisions, such as integration steadiness with the grid, and management of energy storage systems, both of which are of increasing importance in light of the growing scale and sophistication of modern-day offshore wind farms.

\section{Conclusion}
In this study, we performed an exhaustive sensitivity analysis to reveal the impact of atmospheric and marine conditions on the first and second-order properties of offshore wind power. We coupled Gaussian Processes (GPs), a nonparametric statistical learning approach, with OpenFAST, a high-fidelity physics-based simulator, to reproduce, and further predict the design power curve of a 15 MW wind turbine. The analysis revealed that in addition to wind speed, air density and relative wind direction are instrumental in predicting the mean offshore power output, yielding substantial improvements relative to the method of bins. %GP model was shown more accurate than the traditional method of bins and it can be used to analyze the relative importance of the sensitivity factors, which is supported by an analytical analysis. Using the GP regression analysis, we found the order of the sensitivity for the mean power output is wind speed, wind direction, and air density; while the order of the sensitivity of the power output fluctuation is the wind speed, wave direction, and wave height. 
In addition, the impact of wave-related variables on the second-order properties of offshore wind power is, for the first time, elucidated, potentially providing valuable information to power grid integration and storage decisions. OpenFAST is also found suitable to predict the power fluctuation that the traditional power curve cannot capture. Tested on actual offshore measurements from the New York/New Jersey bight, the GP-based models showed significant improvements over the method of bins in predicting the mean and standard deviation of offshore wind power, motivating a need, for both the research community and practitioners in the offshore wind industry, to shift towards multi-input statistical characterizations for offshore wind power assessment and prediction, especially in complex marine environments.
%he results obtained in the case study for the New Jersey offshore area indicate that using GP method with the accuracy of 98.5\% can improve the predictions by up to 6.6\% in comparison with the traditional method of bins with the accuracy of 91.9\%. This difference in the period of only one week is equivalent to approximately 63.4 MWh out of the total operational wind energy of 1347 MWh. 
%This significant difference suggests that the industrial application should consider to shift to the multi-variable statistical model to improve their prediction of power output if a high accuracy is demanded. Insights are also gained to inform the power generation steadiness and compatibility in grid integration.

This study will serve as an anchor point for future research along several directions. For instance, few environmental parameters are missing in the presented modeling and analysis. We plan to fill the gap in the future by extending our analysis to include the effects of turbulence characteristics, vaning, and shearing. In addition, due to the limitation of OpenFAST, the coupling among the dynamics of underwater structure, the turbine, and surface structures is still missing. This may lead to an underestimate of the second-moment dynamics due to waves and related parameters. We expect future updates of OpenFAST to improve this aspect to allow a fully coupled modeling study. At last, it will be beneficial to engage field data to further validate the discovery of this paper and our team is currently exploring opportunities along this direction. 

\section*{Acknowledgement}
This research is partially supported by the Department of Energy (DOE) Advanced Research Projects Agency-Energy (ARPA-E) Program award DE-AR0001186 entitled ``Computationally Efficient Control Co-Design Optimization Framework with Mixed-Fidelity Fluid and Structure Analysis.'' The authors thank DOE ARPA-E Aerodynamic Turbines Lighter and Afloat with Nautical Technologies and Integrated Servo-control (ATLANTIS) Program led by Dr. Mario Garcia-Sanz. The authors acknowledge the Office of Advanced Research Computing (OARC) at Rutgers, The State University of New Jersey for providing access to the Amarel cluster and associated research computing resources that have contributed to the results reported here. URL: https://oarc.rutgers.edu

\appendix
\section*{Appendix A} \label{Appendix A}
\renewcommand{\theequation}{A.\arabic{equation}}

Using the derivatives of the fundamental wind power equation, the significance of its input variables in predicting mean wind power can be estimated given the ranges of the input variables.   
\begin{equation}
    P = \frac{1}{2} C_p \rho A V^3 = \frac{1}{2} C_p \rho A V^3 cos^3(\theta)
    \label{eq:fundamental}
\end{equation}
\begin{equation}
    \frac{\partial P}{\partial \theta} = -\frac{3}{2} C_p \rho A V^3 sin(\theta) cos^2(\theta)
\end{equation}
\begin{equation}
    \frac{\partial P}{\partial \rho} = \frac{1}{2} C_p A V^3 cos^3(\theta)
\end{equation}
The absolute value of the ratio of those derivatives, derived in Eq.~(\ref{eq:rat}), can reveal insights about the relative importance of density and relative wind direction in predicting mean wind power by Eq.~(\ref{eq:fundamental}).
\begin{equation}
    \displaystyle\left\lvert  \frac{\partial P / \partial \theta}{\partial P / \partial \rho} \right\rvert = \frac{\frac{3}{2} C_p \rho A V^3 sin(\theta) cos^2(\theta)} {\frac{1}{2} C_p A V^3 cos^3(\theta)} = 3 \rho \; tan(\theta)
    \label{eq:rat}
\end{equation}

\appendix
\section*{Appendix B} \label{Appendix B}
\renewcommand{\theequation}{B.\arabic{equation}}

 The set of environmental inputs, which is used to explain the variability in the offshore power output $P$, is denoted by $\mathbf{x} \in \mathbb{R}^p$. At our disposal is a set of input-output training data, wherein $[\mathbf{x}_1, ..., \mathbf{x}_n]^T$ is the $n \times p$ matrix of training data inputs, and $\mathbf{P} = P(\mathbf{x}_1, ..., \mathbf{x}_n) =  [P_1, ..., P_n]^T$ is the $n \times 1$ vector of correspondent target values (power mean or standard deviation) collected from the output of the stochastic simulations. The goal of GPs (and of any statistical regression method) is to construct the unknown functional response governing the relationship between $\mathbf{x} \in \mathbb{R}^p$ and $P$ using the training data. GPs can be written in the additive form of Eq.~(\ref{eq:gp1})
\begin{equation}
    P_i = \mu(\mathbf{x}_i) + \xi(\mathbf{x}_i),
    \label{eq:gp1}
\end{equation}
 where $\mu(\mathbf{x}_i)$ is often called \textit{mean function}, while $\xi(\mathbf{x}_i)$ is a dependent error term. The mean function can be modeled either as a constant, or as a parametric function of the inputs $\mathbf{x}$, such that $\mu(\mathbf{x}_i) = \mathbf{x}_i\pmb{\beta}$, where $\mathbf{x}_i$ is the $1 \times p$ vector of regression inputs and $\pmb{\beta}$ is a $p \times 1$ vector of regression coefficients. \par 
 
 The dependent error term $\xi(\mathbf{x}_i)$ is defined as a zero mean Gaussian random field, with an $n \times n$ covariance matrix denoted by $\pmb{\Sigma}$, for which each entry represents a measure of covariance (or similarity) between a pair of data points. The entries of $\pmb{\Sigma}$ can be determined via an isotropic covariance function (or a kernel) $K(\cdot, \cdot)$. Covariance functions play a critical role in GP regression, especially for modeling complex physical processes \cite{ezzat2020turbine}. A popular choice for $K(\cdot, \cdot)$ is the automatic relevance determination squared exponential (ARD-SE) kernel, which is expressed as in Eq.~(\ref{eq:gp2}). 
 \begin{equation}
     K_{\text{ARD-SE}}(\mathbf{x}_i, \mathbf{x}_j) = \sigma_f^2 \exp\bigg[-\frac{1}{2} \sum_{k=1}^p \frac{(\mathbf{x}_{ik} - \mathbf{x}_{jk})^2}{\ell_k^2}\bigg] + \sigma_n^2 \mathbb{I}({i = j}),
     \label{eq:gp2}
 \end{equation}
 where $\mathbf{x}_{ik}$ and $\mathbf{x}_{jk}$ denote the value of the $k$th environmental input for data points $i$ and $j$, respectively. The parameter $\sigma_f^2$ denotes the marginal variance parameter, while $\sigma_n^2$ denotes the noise variance to reflect the stochastic nature of the simulations, such that $\mathbb{I}(\cdot)$ is the indicator function (i.e. $\mathbb{I}(i = j) = 1$ when $i = j$ and $0$ otherwise). The parameters $\ell_1, ..., \ell_p$ are called the length-scale parameters, for which the estimated values can be used to infer the importance of input variables, i.e. smaller values indicate higher importance (or relevance), and hence is the term \textit{automatic relevance determination} in the GP literature \citep{williams2006gaussian}. 
 
 Combined, the regression coefficients of the mean function and the covariance parameters form the vector of GP parameters $\pmb{\Theta} = \{\pmb{\beta}, \sigma_f^2, \ell_1, ..., \ell_p, \sigma_n^2\}$, which can be estimated in a data-driven way by maximizing (minimizing) the (negative) log-likelihood of the GP model. This can be realized numerically using gradient-descent based optimization. Once the set of parameters has been estimated, encapsulated in the vector $\hat{\pmb{\Theta}}$, they form the basis for the GP-based predictions, which can be expressed as in Eq.~(\ref{eq:krig1}). 
 \begin{equation}
     \hat{P}(\mathbf{x}^*) = \mathbb{E}[P(\mathbf{x}^*)|P_1, ..., P_n] = \mu(\mathbf{x}^*) + \hat{\mathbf{k}}^T \hat{\pmb{\Sigma}}^{-1} (\mathbf{P} - \pmb{\mu}),
     \label{eq:krig1}
 \end{equation}
 where $\hat{P}(\mathbf{x}^*)$ is the predicted target value at any arbitrary input point $\mathbf{x}^*$, while $\hat{\pmb{\Sigma}}$ is the estimated covariance matrix using $\hat{\pmb{\Theta}}$. The $n \times 1$ vector $\hat{\mathbf{k}}$ denotes the vector of covariances between $\mathbf{x}^*$ and  $[\mathbf{x}_1, ..., \mathbf{x}_n]^T$, while $\pmb{\mu} = \mu(\mathbf{x}_1, ..., \mathbf{x}_n)$ is the $n \times 1$ vector of mean function evaluations at the training data. \par
 
 For the analysis in this paper, we have implemented our own version of the GP regression model. %in the statistical programming language $\texttt{R}$
 For parameter estimation, the negative log-likelihood was minimized using the gradient descent-based minimization % command $\texttt{nlm}$ in $\texttt{R}$ 
 with an iteration limit set at $300$. Setting $\mu(\mathbf{x})$ to zero appeared to return satisfactory predictive performance for all the GP models considered herein.

\bibliography{mybibfile}

\end{document}